%% file: main.tex
\documentclass[a4paper,11pt]{article}

\usepackage{jinstpub} 
\usepackage{tabulary}
\usepackage{placeins} 
\usepackage{amssymb}
\usepackage{lineno}
\usepackage{caption}
\usepackage{subcaption}
\usepackage{amsmath}
\usepackage{amsthm} 
\usepackage{bm} 
\usepackage{xspace}
\usepackage{upgreek}
\usepackage{float}
\usepackage{booktabs}  
\usepackage[svgnames]{xcolor}

\newcolumntype{K}[1]{>{\centering\arraybackslash}p{#1}}

\newcommand{\gf}{\textsc{Geant}4\xspace}
\newcommand{\minerva}{MINERvA\xspace}
\newcommand{\genie}{GENIE\xspace}
\newcommand{\uboone}{MicroBooNE\xspace}

\title{Neutral pion reconstruction using machine learning in the \minerva experiment at $\langle E_\nu \rangle \sim 6$ GeV }

\newcommand{\Rutgers}{Rutgers, The State University of New Jersey, Piscataway, New Jersey 08854, USA}

\newcommand{\Florida}{University of Florida, Department of Physics, Gainesville, FL 32611}

\newcommand{\CBPF}{Centro Brasileiro de Pesquisas F\'{i}sicas, Rua Dr. Xavier Sigaud 150, Urca, Rio de Janeiro, Rio de Janeiro, 22290-180, Brazil}
\newcommand{\PUCP}{Secci\'{o}n F\'{i}sica, Departamento de Ciencias, Pontificia Universidad Cat\'{o}lica del Per\'{u}, Apartado 1761, Lima, Per\'{u}}

\newcommand{\Pittsburgh}{Department of Physics and Astronomy, University of Pittsburgh, Pittsburgh, Pennsylvania 15260, USA}
\newcommand{\Guanajuato}{Campus Le\'{o}n y Campus Guanajuato, Universidad de Guanajuato, Lascurain de Retana No. 5, Colonia Centro, Guanajuato 36000, Guanajuato M\'{e}xico.}

\newcommand{\Tufts}{Physics Department, Tufts University, Medford, Massachusetts 02155, USA}
\newcommand{\WM}{Department of Physics, College of William \& Mary, Williamsburg, Virginia 23187, USA}
\newcommand{\FNAL}{Fermi National Accelerator Laboratory, Batavia, Illinois 60510, USA}

\newcommand{\MCLA}{Massachusetts College of Liberal Arts, 375 Church Street, North Adams, MA 01247}
\newcommand{\UMD}{Department of Physics, University of Minnesota -- Duluth, Duluth, Minnesota 55812, USA}

\newcommand{\UNI}{Facultad de Ciencias, Universidad Nacional de Ingenier\'{i}a, Apartado 31139, Lima, Per\'{u}}
\newcommand{\Rochester}{University of Rochester, Rochester, New York 14627 USA}

\newcommand{\USM}{Departamento de F\'{i}sica, Universidad T\'{e}cnica Federico Santa Mar\'{i}a, Avenida Espa\~{n}a 1680 Casilla 110-V, Valpara\'{i}so, Chile}

\newcommand{\OregonState}{Department of Physics, Oregon State University, Corvallis, Oregon 97331, USA}
\newcommand{\oxford}{Oxford University, Department of Physics, Oxford, OX1 3PJ United Kingdom}

\newcommand{\upenn}{Department of Physics and Astronomy, University of Pennsylvania, Philadelphia, PA 19104}
\newcommand{\AMU}{AMU Campus, Aligarh, Uttar Pradesh 202001, India}

\newcommand{\Mohali}{Department of Physical Sciences, IISER Mohali, Knowledge City, SAS Nagar, Mohali - 140306, Punjab, India}
\newcommand{\CINVESTAV}{Departamento de Fisica Col. San Pedro Zacatenco, 07360 Mexico, DF, Av. Centro de Investigación y de Estudios Avanzados del Instituto Politecnico Nacional (Cinvestav), Mexico City, Mexico}
\newcommand{\york}{York University, Department of Physics and Astronomy, Toronto, Ontario, M3J 1P3 Canada}

\newcommand{\nUSM}{a}
\newcommand{\nCBPF}{b}
\newcommand{\nAMU}{c}
\newcommand{\nPUCP}{d}
\newcommand{\nOregonState}{e}
\newcommand{\nRochester}{f}
\newcommand{\nGuanajuato}{g}
\newcommand{\nWM}{h}
\newcommand{\nFNAL}{i}
\newcommand{\nUMD}{j}
\newcommand{\nyork}{k}
\newcommand{\nMohali}{l}
\newcommand{\nupenn}{m}
\newcommand{\nTufts}{n}
\newcommand{\nRutgers}{o}
\newcommand{\noxford}{p}
\newcommand{\nMCLA}{q}
\newcommand{\nCINVESTAV}{s}
\newcommand{\nPittsburgh}{r}
\newcommand{\nFlorida}{t}
\newcommand{\nUNI}{u}
  \author[\nUSM,\nCBPF,1] {A.~Ghosh\note{Corresponding author, now at University of Padova.}}
  \author[\nUSM]          {B.~Yaeggy}
  \author[\nUSM]          {R.Galindo}
  \author[\nAMU]       {Z.~~Ahmad~Dar}
  \author[\nAMU]          {F.~Akbar}
  \author[\nPUCP]         {M.~V.~Ascencio}
  \author[\nOregonState]  {A.~Bashyal}
  \author[\nRochester]    {A.~Bercellie}
  \author[\nGuanajuato]   {J.~L.~Bonilla}
  \author[\nCBPF]         {G.~Caceres}
  \author[\nRochester]{T.~Cai}
   \author[\nOregonState,\nCBPF]{M.F.~Carneiro}
  \author[\nCBPF]         {H.~da~Motta}
  \author[\nRochester,\nPUCP]{G.A.~D\'{i}az~}
  \author[\nGuanajuato]   {J.~Felix}
  \author[\nWM]           {A.~Filkins}
  \author[\nRochester]    {R.~Fine}
  \author[\nPUCP]         {A.M.~Gago}
  \author[\nFNAL,\nRochester]{T. Golan,}
  \author[\nUMD]          {R.~Gran}
  \author[\nFNAL,\nyork]    {D.A.~Harris}
  \author[\nRochester]    {S.~Henry}
  \author[\nMohali]       {S.~Jena}
  \author[\nFNAL]         {D.~Jena}
  \author[\nRochester]    {J.~Kleykamp}
  \author[\nWM]           {M.~Kordosky}
  \author[\nupenn]        {D.~Last}
  \author[\nTufts,\nRutgers]{T.~Le}
  \author[\nCBPF]         {A.~Lozano}
  \author[\noxford]       {X.-G.~Lu}
   \author[\nMCLA]         {E.~Maher} 
  \author[\nRochester]    {S.~Manly}
  \author[\nTufts]        {W.A.~Mann}
  \author[\nupenn]        {C.~Mauger}
  \author[\nRochester]    {K.S.~McFarland}
  \author[\nPittsburgh]{B.~Messerly}
  \author[\nUSM]          {J.~Miller}
  \author[\nCINVESTAV]    {Luis~M.~Montano}
  \author[\nPittsburgh]   {D.~Naples}
  \author[\nWM]{J.K.~Nelson}
  \author[\nFlorida]      {C.~Nguyen}
  \author[\nRochester]    {A.~Olivier}
  \author[\nPittsburgh]   {V.~Paolone}
  \author[\nFNAL,\nRochester]{G.N.~Perdue}
  \author[\nupenn,\nGuanajuato]{M.A.~Ram\'{i}rez}
  \author[\nFlorida]      {H.~Ray}
  \author[\nRochester]{D.~Ruterbories}
  \author[\nUNI]          {C.J.~Solano~Salinas}
  \author[\nPittsburgh]   {H.~Su}
  \author[\nRochester]    {M.~Sultana}
  \author[\nTufts]        {V.S.~Syrotenko}
  \author[\nWM,\nGuanajuato]{E.~Valencia}
  \author[\nFlorida]      {M.Wospakrik}
  \author[\nRochester]    {C.~Wret}
  \author[\noxford]       {K.~Yang}
  \author[\nWM]           {L.~Zazueta}
  \affiliation[\nWM]{\WM}
  \affiliation[\nAMU]{\AMU}
  \affiliation[\nPUCP]{\PUCP}
  \affiliation[\nOregonState]{\OregonState}
  \affiliation[\nRochester]{\Rochester}
  \affiliation[\nGuanajuato]{\Guanajuato}
  \affiliation[\nCBPF]{\CBPF}
  \affiliation[\nUSM]{\USM}
  \affiliation[\nUMD]{\UMD}
  \affiliation[\nyork]{\york}
  \affiliation[\nFNAL]{\FNAL}
  \affiliation[\nMohali]{\Mohali}
  \affiliation[\nupenn]{\upenn}
  \affiliation[\nTufts]{\Tufts}
  \affiliation[\nRutgers]{\Rutgers}
  \affiliation[\noxford]{\oxford}
  \affiliation[\nMCLA]{\MCLA}
  \affiliation[\nCINVESTAV]{\CINVESTAV}
  \affiliation[\nPittsburgh]{\Pittsburgh}
  \affiliation[\nFlorida]{\Florida}
  \affiliation[\nUNI]{\UNI}

\emailAdd{anushree.ghosh@unipd.it}
\emailAdd{millerjonathanandrew@gmail.com}

\abstract{
This paper presents a novel neutral-pion reconstruction that takes advantage of the machine learning technique of semantic segmentation using MINERvA data collected between 2013-2017, with an average neutrino energy of $6$ GeV.
Semantic segmentation improves the purity of neutral pion reconstruction from two $\gamma$s from $70.7 \pm 0.9$\% to $89.3 \pm 0.7$\% 
and improves the efficiency of the reconstruction by approximately 40\%.
We demonstrate our method in a charged current neutral pion production analysis where a single neutral pion is reconstructed.
This technique is applicable to modern tracking calorimeters, such as the new generation of liquid-argon time projection chambers, exposed to neutrino beams with $\langle E_\nu \rangle$ between 1-10 GeV.
In such experiments it can facilitate the identification of ionization hits which are associated with electromagnetic showers, thereby enabling improved reconstruction of charged-current $\nu_e$ events arising from $\nu_{\mu} \rightarrow \nu_{e}$ appearance.

}


\arxivnumber{XXXX.YYYYY} 

\begin{document}
\maketitle
\flushbottom

\section{Introduction}
\label{sec:intro}

\minerva's extensive data set is used to investigate a wide variety of neutrino interaction channels, including both neutral and charged pion production.
Neutral pion production is challenging to measure because the neutral pion reconstruction relies on identifying two photon-like objects in an event, and until now the primary approach to reduce backgrounds has been to simply cut on the neutral pion invariant mass.
This is particularly true for the \minerva dataset taken between 2013-17, called NuMI \cite{Adamson:2015dkw} Medium Energy Beam data, where most neutral pions are produced with a kinetic energy between 0.015 - 2.0 GeV at $\langle E_\nu \rangle \sim 6$ GeV.
This dataset can be contrasted with the dataset for 2009-2012, called Low Energy Beam data, where most neutral pions are produced with a kinetic energy between 0.015 - 1.0 GeV at $\langle E_\nu \rangle \sim 3.5$ GeV.  
The increased number and energy of particles produced by the neutrino interaction at $\langle E_\nu \rangle \sim 6$ GeV creates a complicated environment for neutral pion reconstruction.
For the results presented here, we use a subset of the Medium Energy Beam data corresponding to $1.65\times10^{20}$ protons on target.  

Machine Learning (ML) reconstruction applications traditionally have taken one of two approaches \cite{Psihas:2020pby}\cite{Guest_2018}\cite{Qasim_2019}\cite{Andrews_2018}\cite{ad766625192f4650abd255078010c7cb}.
In the first approach, the entire record from an event is provided to the machine learning algorithm (MLA) \cite{Aurisano_2016}\cite{Adamson_2016}.
This may prove problematic for exclusive or semi-exclusive measurements.
In such measurements, many or most particles produced by the interaction are reconstructed.
A challenge for this technique is the possibility that the different algorithms may claim conflicting hypotheses for the same information.  In the second approach, which grew out of the log-likelihood based reconstructions \cite{Acero_2020}, \cite{Eberly_2015}, only a subset of the data is provided to the MLA.
Here the problem is no longer that information may be used to make incompatible hypotheses, but rather that valuable information may not be provided to the algorithms.

Deep Convolutional Neural Networks \cite{lecun1995comparison}\cite{DBLP:journals/corr/RazavianASC14} (DCNN) using techniques like semantic segmentation \cite{DBLP:journals/corr/LongSD14} make it possible to identify individual objects in an input image where pixels in the image are related by some regular metric (in \cite{Qasim_2019} a distance weighted graph network is used for irregular geometry).
These developments could revolutionize the reconstruction of events in some high energy particle physic experiments, in cases where the data can be represented by images with regular geometry, by allowing DCNN methods to model the activity in the detector as a step in the characterization of the interaction.
This improves the measurement of energy that is deposited in the detector and motivates a third approach, where MLA provides context to the reconstruction \cite{PhysRevD.100.073005}.

In semantic segmentation, the MLA classifies individual bits of information (ionization hits in \minerva) as one of several interesting classes.
The classification then can be used as a filter for the hits used in the reconstruction algorithms (in \cite{DiFlorio:2018res} a CNN was used to construct a hit filter for track pixel seeds used as input to the CMS track reconstruction algorithm ).
The hits and information can be provided to further
MLAs, however the clearest way to test the impact of the MLA responsible for semantic segmentation
is to use it as a filter for the current state-of-the-art algorithm.

The ability to identify electromagnetic-like energy depositions within complicated event topologies is also important for the identification of electrons in charged-current electron-neutrino interactions.
Improving the identification of such showers would improve both the efficiency and purity of measurements of electron neutrino appearance in current and future experiments, which are a crucial component of the worldwide effort to measure neutrino oscillations.

\subsection{Overview of the paper}
\label{sec:paperplan}

The paper is organized as follows. 
First, Section \ref{sec:background} provides a short review of details of the \minerva detector, details of the NuMI Medium Energy Beam data set and neutral pion production analyses in \minerva.
Section \ref{sec:pi0_rec_ML} describes the development of a machine learning algorithm for the semantic segmentation of energy deposition within the detector based on the class of particle which deposited that energy.
Section \ref{sec:stateofartl} discusses the neutral pion reconstruction algorithm and an extension using semantic segmentation as a filter of electromagnetic-like hits.
Section \ref{sec:conclusions} presents conclusions and discusses implications for other experiments.

\input{sec_background}

\input{pi0_Reconstruction_Semantic_Segmentation}

\input{sec_tradanalysis}
\input{sec_conclusion}
\input{acknowledgments}



\appendix
\input{sec_discussion}

\bibliographystyle{JHEP}
\bibliography{main}


\end{document}

%% file: sec_background.tex
\section{Background}
\label{sec:background}
The main goal of the \minerva experiment is to measure neutrino-nucleus interaction cross-sections with a high degree of precision.
Resonance production is the most abundant interaction mechanism at energies relevant to DUNE \cite{Abi:2018dnh}, a flagship neutrino experiment using a neutrino beam whose energy is comparable to that of \minerva.
Thus, measurements of baryon resonance production yielding final state pions are crucial.

In this work we study $\pi^0$ reconstruction algorithms applied to neutrino interactions on hydrocarbon (CH) target using the NuMI Medium Energy Beam dataset.
To study the performance of these algorithms, we use a sample of candidates in the semi-exclusive channel $\nu_{\mu} + CH \longrightarrow \mu^{-} + \pi^{0} +$ \textit{nucleons}.  Performance in an alternative  semi-inclusive channel, where instead of just nucleons anything is allowed other than a neutral pion, is considered in Appendix~\ref{sec:semiinclusive}.




\subsection{The \minerva Experiment}
\label{sec:minerva}

\begin{figure}[!htb]
\centering
\begin{subfigure}{0.5\textwidth}
  \centering
  \includegraphics[width=1.0\linewidth]{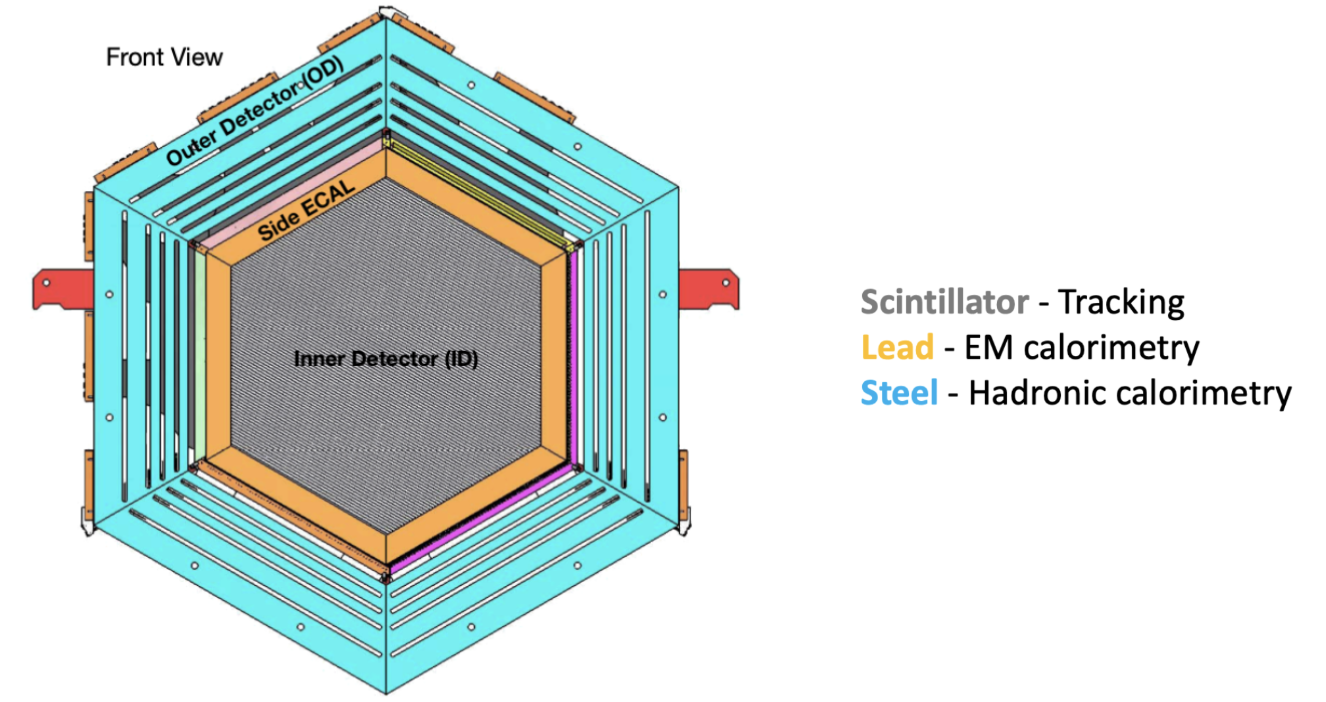}
  \caption{Front View - Inner Detector}
  \label{fig:frontView2}
\end{subfigure}%
\begin{subfigure}{.5\textwidth}
  \centering
  \includegraphics[width=1.0\linewidth]{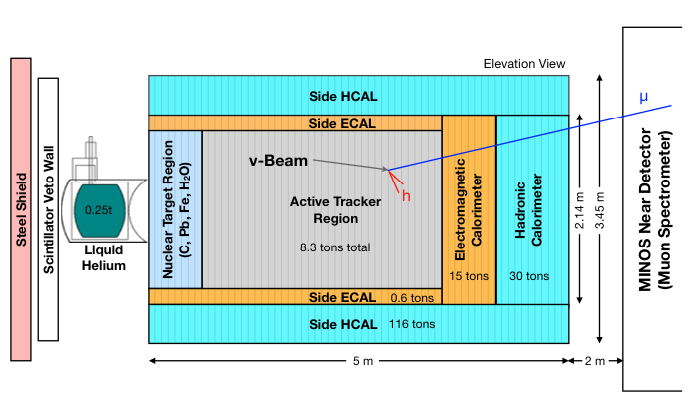}
  \caption{Side View - Detector}
  \label{fig:nuclearT}
\end{subfigure}
\caption[The \minerva Detector]{a) The inner detector has four regions, which are the nuclear target region, active tracking region (tracker), downstream electromagnetic calorimeter (ECAL) and downstream hadronic calorimeter (HCAL) region. b) 
Side view of the \minerva detector with highlighted regions: nuclear targets, active tracker, calorimeters and MINOS near detector.
The nuclear target region consists of passive material (Pb, Fe and C) interspersed with scintillator.
From \cite{Aliaga:2013uqz}.
}
\label{fig:detector}
\end{figure}

Neutrinos are produced by decays of charged pions produced by 120 GeV protons incident on a fixed carbon target and focused with two magnetic horns to create the NuMI beam.
The muons associated with those same decays are stopped by approximately 200 meters of rock, leaving a beam of neutrinos.
The \minerva detector is located directly in front of the MINOS near detector \cite{Perdue:2012hg}\cite{Michael:2008bc}.
The core of the detector is a volume with a hexagonal cross section with 120 modules along the beam direction ($z$), 5m long, and 1.7m of apothem, as shown in Fig. \ref{fig:detector}.
Starting from upstream and proceeding downstream, the full detector consists of multiple nuclear targets (He, Pb, Fe, C, H$_2$O), a central scintillator tracker inner detector, an electromagnetic calorimeter (ECAL), a hadronic calorimeter (HCAL), and the MINOS near detector, which serves as a magnetized iron spectrometer to identify the charge and momentum of muons. 

The inner detector is made up of 8.3 tons of polystyrene (CH) strips divided into planes in three orientations: X, U and V (U and V are offset by $\pm 60$ degrees from the orientation of the X-view).
These orientations allow the planes to be divided into three classes defining ``views'', the X-view, U-view and V-view.
Each view can also be represented as a two dimensional image, i.e. the X-view can be represented as a x-z image.
The planes are ordered in the pattern UXVXUXVX such that the sampling frequency of the X-view is double that of the U-view and V-view along the z-axis. 
Further details are available in \cite{Aliaga:2013uqz}\cite{Perdue:2012hg}.

\subsection{Neutrino-nucleus interactions}
\label{sec:nuint}

In pursuit of a suite of cross-section measurements covering $\nu_e/\bar{\nu}_e$ and $\nu_{\mu}/\bar{\nu}_{\mu}$ interactions \cite{Fields_2013}\cite{Fiorentini_2013}
\cite{Tice_2014}\cite{Wolcott_2016}\cite{Park_2016}\cite{Ruterbories_2019}\cite{Carneiro_2020}, \minerva has made a series of resonance production measurements including of neutral pion production in the NuMI Low Energy Beam dataset, $\langle E_\nu \rangle \sim 3.5$ GeV.
Extending these measurements to the NuMI Medium Energy Beam dataset, $\langle E_\nu \rangle \sim 6$ GeV,  provides an opportunity to expand the phase space of the earlier measurement, but also introduces challenges given the higher intensity and higher mean energy of the Medium Energy beam.

A neutrino interaction can produce a pion through several different processes.
One process proceeds through baryon resonance production, in which the neutrino excites the target nucleon to a baryon resonance state that then decays to a nucleon and a pion in $\sim 10^{-23}$ seconds.
Another process is deep inelastic scattering (DIS), in which the neutrino interacts with one of the quarks inside the nucleon and the quark then hadronizes to form one or multiple pions.
The main characteristic of DIS events is that the struck quark is ejected from the nucleon as opposed to being merely transformed within the nucleon, unlike other pion production processes.
Pion production from a nucleon not involving a resonance is commonly called a non-resonant process and it is treated as a sub-sample of DIS processes.
A third process is coherent pion production from nuclei, a process in which a pion is created from neutrino-nucleus interaction where the target nucleus remains unchanged in its ground state after the scattering.

\subsection{Neutral pion production in \minerva}
\label{sec:analysis}

\begin{figure}[!htb]
\centering
  \includegraphics[height=4.7cm, width = 15.0cm]{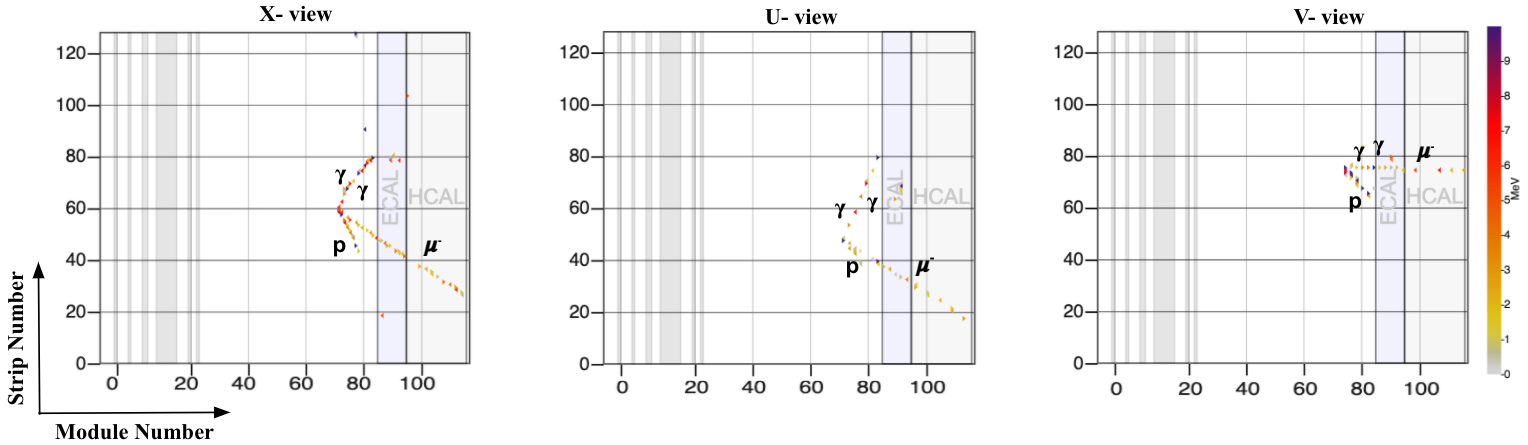}
  \caption{Arachne display of a simulated DIS event with a single neutral pion in the final state. 
  Arachne is the \minerva event viewer and is useful for developing and analyzing reconstructions, evaluating data quality during the \minerva's data run, and providing a natural image representation of the \minerva data, \cite{Tagg:2011wk}.
  }
  \label{fig:BarbSignal_ArachneDisplay}
\end{figure}

In the semi-exclusive neutral pion production analyses, the signal is $\pi^0 + X + \mu$ where $X$ can not be a meson.
An example of this analysis is given by \cite{Altinok:2017xua}.
This type of analysis provides the central context for our  application and the discussion in Section \ref{sec:stateofartl}.
Also examined in this work is a less selective analysis, here referenced as semi-inclusive with figures and details in Appendix \ref{sec:semiinclusive}, which is defined by a signal $\pi^0 + X + \mu$ where $X$ is not a neutral pion.
An example of a neutral pion production event relevant for semi-exclusive analyses is provided in Fig. \ref{fig:BarbSignal_ArachneDisplay}.

Other charged current neutral pion production analyses, not examined in this work, are  variations of the measurements described above.
A more selective analysis is defined by a signal $\mu^- + \pi^0 + P + X$ where $X$ can not be a meson or any energetic proton.
An example of such an analysis is given by \cite{Coplowe:2020yea}.
In other analyses $X$ is minimal, such as in anti-neutrino analyses \cite{Aliaga:2015wva} and in analyses of neutral current neutral pion production \cite{Wolcott:2016hws}.

\subsection{\genie and \gf simulation in \minerva}
\label{sec:simulation}
Detailed event simulation is essential for the development of  reconstruction algorithms and analysis of MINERvA data.
We use an event generator to simulate neutrino-nucleus interactions and a detector simulation to simulate the response of the MINERvA detector to the products of those interactions.
Our event generator is \genie v2.12.6 \cite{genie1}, and our detector simulation package is based on \gf v9.4.2 \cite{Agostinelli:2002hh}.
\genie models neutrino interactions, while the
the final state particles generated by GENIE are propagated through the \minerva detector by \gf.

The charge current (CC) pion production processes relevant to this work proceed via baryon resonance production or by DIS. For the baryon resonance channels, 
\genie uses the Rein and Sehgal model without interference between neighboring resonances \cite{REIN198179}. It models DIS using the the Bodek and Yang model \cite{bodek2003higher}.
The nuclear medium is simulated by GENIE based on a modified version of the relativistic Fermi gas model that includes short-range nucleon-nucleon correlations \cite{PhysRevD.24.1400}. 
GENIE models final-state interactions of pions and nucleons (FSI) within the primary nucleus using an effective cascade model~\citep{doi:10.1063/1.3661588}.

\genie output for the final state particle kinematics after FSI is used as the input for \gf, which propagates particles in discrete steps.
In each step, the energy deposited in the detector is calculated due to ionization and radiation based on the particle type.
After each step, \gf uses interaction models to determine whether any of the particles interacted in the detector.
If an interaction occurs, \gf produces outgoing particles according to the interaction model.

%% file: pi0_Reconstruction_Semantic_Segmentation.tex
\section{Semantic segmentation of energy deposition within \minerva}
\label{sec:pi0_rec_ML}

Our machine learning algorithm was developed to provide context for further reconstructions in \minerva, in particular the neutral pion reconstruction.
We built a fully connected network (FCN) \cite{DBLP:journals/corr/LongSD14}  to realize a semantic segmentation of the hits within the views of \minerva.
This provides additional information for each hit to physics reconstruction algorithms in \minerva, such as the neutral pion reconstruction algorithm.
While this machine learning algorithm was developed with the goal of providing context to the neutral pion reconstruction, it was developed in an analysis independent way, only assuming that the neutrino interaction was a charged current interaction where a muon was measured in the MINOS spectrometer.
Alternative machine learning models are described Appendix \ref{sec:alternativesegmentations}.
These includes ones where the requirement of a regular geometry is bent by including as additional layers the other \minerva views, which are spatially shifted and which require extrapolation relative to the X-view.


\subsection{Data description}
\label{sec:datadescription}

The dataset used to train this MLA was collected as described in \cite{Perdue:2018ihs} but with an additional label.
In \cite{Perdue:2018ihs} the data was simulated with flux from the Medium Energy configuration of the NuMI beam.
The simulation has been tuned based on previous \minerva analyses (for a recent description, \cite{Stowell_2019}) but each analysis using sides bands for a final correction as a final step which has not been done here during the development of the reconstruction.
The data shown represents approximately 10\% of the total data available to the semi-exclusive neutral pion production analysis, and the MC statistics shown is about twice that of the data.
The MC used is semi-inclusive and includes the target region.

The additional label vector was the full image but with each pixel labeled as ``electromagnetic(EM)'', ``neutron'', ``non-electromagnetic(non-EM)'', or ``null''.
This label looks up the history of the highest energy deposited in that pixel and selects the source that produced that energy deposit, which is not necessarily the particle that was produced directly in the neutrino interaction.
As such the label defines the source energy which was deposited in the detector and is not necessarily directly related to particles that were produced in the neutrino interaction.
If the immediate history was a $\gamma$, $e$ or $\pi^0$, the pixel was defined as being from the ``EM'' class.
If the immediate history was a neutron, the pixel was defined as being from the ``neutron'' class.
The remaining pixels were defined as ``non-EM'' unless there was no immediate history in which case they were classed as ``null''.
This ``null'' class includes activity in the detector unrelated to the neutrino interaction.
The majority of the ``null'' class pixels have no energy in the pixel, but some pixels assigned the ``null'' class may have energy due to cross-talk or other backgrounds (including activity in the detector from cosmic rays or residual activity from earlier neutrino interactions).
Each pixel is assigned one and only one class.


\subsection{Network description}
\label{sec:netwrokdescription}

In order to reconstruct neutral pions via the semantic segmentation approach, we labeled each pixel of the image as ``EM'', ``Neutron''  or  ``non-EM''.
The pixel was left without a label (NULL) if it was not given one of these three labels.
This ML task can be described as multi-class classification with 3 classes.
The energy originating from neutrons was explicitly labeled due to the possibility that hits from the neutron could be mistaken for a low energy $\gamma$ shower and to provide the opportunity to reconstruct neutrons at a later stage.
We differentiate the pixel labels from the simulation given in \ref{sec:datadescription} from those that the classifier returns (EM, Neutron, non-EM and NULL) to emphasize that the classifier has 3 classes and not 4.

\begin{figure}[!htb]
\centering
\includegraphics[height= 16cm, width= 13cm]{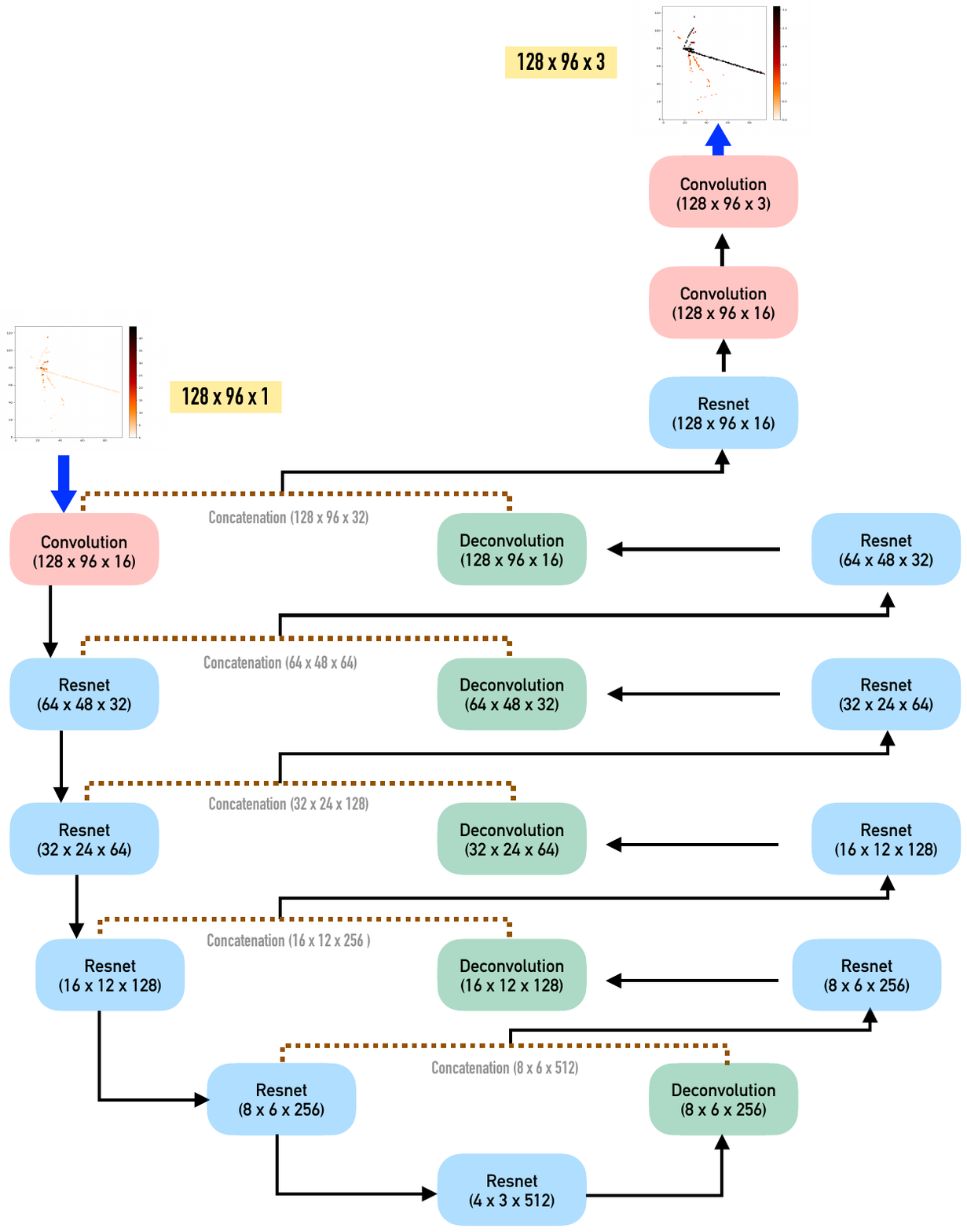}
  \caption{UResNet model architecture. The input image (left) is encoded using ResNet modules along the left before decoding on the right resulting in the semantic segmented  image (top right). During decoding, the output of the encoding ResNet modules is combined at each decoding step. For each step, the width X height X (semantic) depth is provided. This diagram is for the energy only x-view with 4 class output.}
  \label{fig:uresnet}
\end{figure}

The \minerva events are sparse, i.e. the majority of pixels in any given event, $(\sim 99\%)$, are not assigned and are thus left unlabeled.
The package LarCV \cite{larcv} was adopted to save the views as an image in the ROOT \cite{Brun:1997pa} file format and to facilitate the use of the U-ResNet network \cite{uresnetcite} \cite{Adams:2018bvi}.
The implementation of the U-ResNet model used is described by Fig. \ref{fig:uresnet}.
The U-ResNet model was initially developed within the MicroBooNE collaboration \cite{microboone}, and combines U-Net \cite{DBLP:journals/corr/RonnebergerFB15} with ResNet \cite{DBLP:journals/corr/HeZRS15} to create a fully convolutional network for image or particle physics event segmentation.
The skip connections of ResNet allow deeper networks with more expressive power, or the breadth of the domain of functions that the network can compute, to be trained without running into the problem of vanishing gradients during back-propagation based machine learning \cite{726791}.

Each ResNet module consists of two convolution layers and each of these layers is followed by a batch normalization operation~\cite{DBLP:journals/corr/IoffeS15} and a rectified-linear unit (ReLU) activation function.
In Fig. \ref{fig:uresnet}, black arrows indicate the direction of tensor data flow and the brown dashed lines indicate concatenation operations to combine the output of convolution layers from the encoding path to the decoding path.
The final output has the same spatial dimension as the input with a depth defined by the number of labels. 

As described in Section \ref{sec:background}, the data can be represented in three views: X, U and V, which may be interpreted as images.
Each image has two channels: hit time and energy.
Three separate semantic segmentation models were trained using the same algorithm with prediction applied to each view separately.

The detailed description of the production of images containing hit and energy information at the detector can be found in \cite{Perdue:2018ihs}.
For the training of the whole detector configuration, both nuclear target and tracker regions were used.
But for the semantic prediction and the neutral pion reconstruction, only the tracker region was used.

\begin{figure}[!htb]
  \centering
  \includegraphics[height = 7.1cm, width=9.0cm]{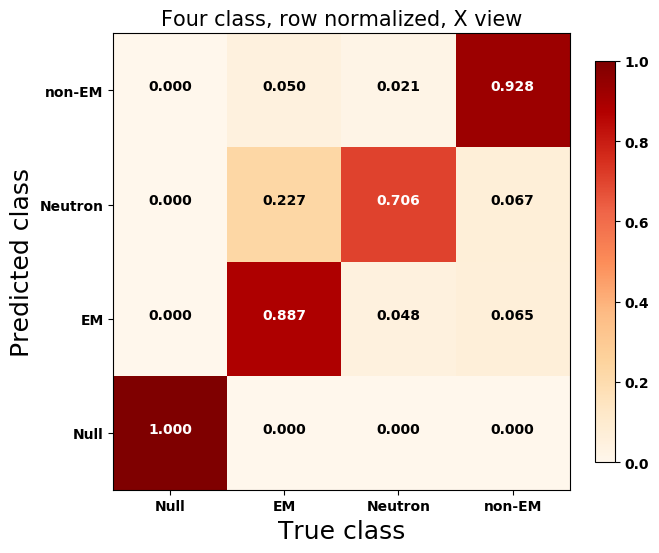}
\caption{Row normalized confusion matrix with four labels: ``null", EM, ``neutron'', non-EM.
Only X-views are used. This confusion matrix is for a sample representative of the full training sample, including the nuclear target region.}
\label{fig:row_norm4}
\end{figure}

Figure \ref{fig:row_norm4} shows the confusion matrix obtained by training over one million simulated Monte Carlo (MC) events and validating over fifty thousand simulated events.
The validation set is independent of the training sample but comes from the same \minerva MC dataset.
A confusion matrix represents the correlation between reconstructed and true values.
The labels classified correctly according to the simulation are represented by the diagonal value, whereas miss-assignments are represented by the off-diagonal cells.
Figure \ref{fig:row_norm4} represents the row-normalized matrix, which can be interpreted as the fraction of events reconstructed under a label (row value) that matches the one that truly originated the event.
The values along the row-normalized matrix diagonal are above 70, which indicates the goodness of the network's ability to predict the data.


We applied semantic segmentation to the X-view only.
We made this decision based on numerous factors, including the dominance of the X-view in the baseline reconstruction, the transparency of the impact of the application of the semantic segmentation to the neutral pion reconstruction and the requirement of a regular geometry for the use of a FCN for semantic segmentation.
In the reconstruction of the neutral pion the X-view is reconstructed first and the sampling of the X-view is twice that of the other views, increasing the importance of the X-view for the reconstruction.
Details about an alternative where we apply semantic segmentation in all three views is provided in Appendix \ref{sec:alternativesegmentations}.
The neutral pion reconstruction would need to be modified to include application to the U and V views, complicating the comparison and evaluation.

 
 Figure \ref{fig:ArachneDisplay_ML} shows the comparison between true MC events and their  corresponding predictions from the MLA.
 The left panel of the Fig. \ref{fig:ArachneDisplay_ML} is a visualization of MC events using a web-based event viewer called Arachne \cite{Tagg:2011wk}. The right panel of the Fig. \ref{fig:ArachneDisplay_ML} is the corresponding predicted hits of the event.
 The colors of the pixel signify the different labels where 0, 1, 2 and 3 represent label null, EM-like, neutron and non-EM-like, respectively.


%% file: sec_tradanalysis.tex
\section{Neutral pion reconstruction}
\label{sec:stateofartl}


Here we present the baseline neutral pion reconstruction algorithm in the context of the semi-exclusive analysis and compare it to the machine learning-based algorithm in a step-by-step manner.
The baseline neutral pion reconstruction algorithm in \minerva is based on the identification of two electromagnetic shower candidates as $\gamma$ candidates produced via the decay $\pi^0 \rightarrow \gamma \gamma$.
The neutral pion reconstruction algorithm is designed to identify and measure the energy and direction of the two electromagnetic candidate showers; the algorithm proceeds in four stages.

The first stage is a hit filter, which removes hits which are used by the muon or proton reconstructions, considered likely to be from optical or electronic cross-talk, or are in the HCAL.
The remaining hits are available to the second stage, the electromagnetic shower candidate reconstruction.
The only change for the machine learning-based algorithm is to modify this hit filter to remove hits which are determined by the machine learning algorithm to be not arising from electromagnetic activity.

In the second stage, the hits are grouped into candidate showers under the hypothesis that they originated from the electromagnetic cascade of an energetic $\gamma$ originating in the neutrino interaction vertex.

While the neutral pion analyses primarily focus on the reconstruction of neutral pions in neutrino interactions with only a single neutral pion (and therefore two electromagnetic showers), reconstructions of one or three electromagnetic shower candidates are also kept.
Three electromagnetic candidate reconstructions are kept due to the likelihood that one of these cannot be reconstructed into a $\gamma$ candidate.
Single electromagnetic candidate reconstructions are kept because the single electromagnetic candidate can possibly be divided into two overlapping showers.

The electromagnetic shower candidates are then available for the third stage, the $\gamma$ candidate reconstruction.
As part of the requirement to be able to be reconstructed as a $\gamma$, only shower candidates comprised of hits from two or three views are considered.

In the final stage, the reconstruction of the $\gamma$ candidates from the electromagnetic candidates is realized.
Here, properties of energetic $\gamma$s are used, such as the conversion length in the scintillator and the average energy per unit length deposited by the shower candidate, $\frac{dE}{dX}$.
Finally, the neutral pion candidate is reconstructed from the $\gamma$ candidates using the invariant mass, $m_{\gamma\gamma}=\sqrt{2 E_{\gamma1} E_{\gamma2} (1-\cos\theta)}$ where $E_{\gamma1}$  and $E_{\gamma2}$ are the energy for $\gamma$ candidate 1 (most energetic)  and for $\gamma$ candidate 2 (less energetic) respectively, $\theta$ is the angle between the $\gamma$ candidates.

\begin{figure}[!htb]
\centering
  \includegraphics[height=7.0cm, width = 15cm]{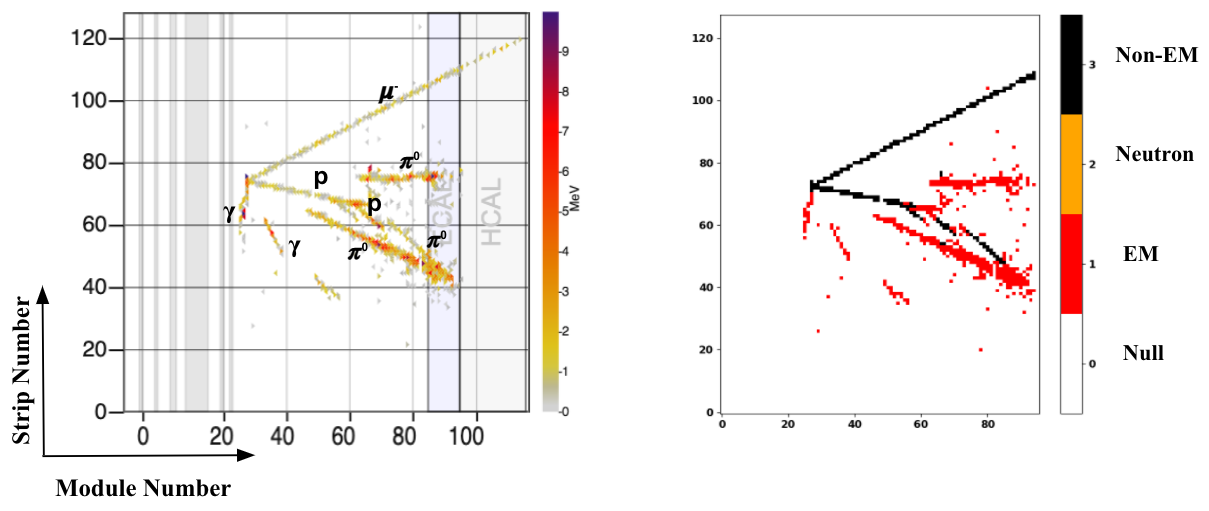}
  \caption{Left: A visualization of a simulated DIS neutral pion production event using the  web-based tool Arachne\cite{Tagg:2011wk} tool. Right: a semantically-segmented image of the same event.  For the medium energy dataset, many events have $\gamma$s with trajectories very close to each other, thereby increasing the difficulty of distinguishing $\gamma$ pairs. The higher the $\pi^{0}$ energy is, the smaller the opening angle is, raising the complexity of distinguishing the $\gamma$s' pair. A common approach when faced with this type of issue is by rejecting these mis-reconstructed events, which reduces the efficiency of $pi^0$ reconstruction. }
  \label{fig:ArachneDisplay_ML}
\end{figure}

Each of the four stages is described in more detail below.  Figure \ref{fig:ArachneDisplay_ML} shows an event that illustrates the challenge to $\gamma$ and neutral pion production in multi-GeV CC inclusive interactions.

The procedure that we use for the application of semantic segmentation to the reconstruction of a neutral pion is to add an additional step to the hit selection, as described below in Sec.\ref{sec:hitselection}.
Having such a simple and obvious application of ML to an existing reconstruction has distinct advantages, including the advantage of transparency when assigning and propagating systematic uncertainties.
The simplicity and clearness of the application can be contrasted with the expressive complexity and opacity of the semantic segmentation model or even the relative algorithmic complexity of the $\Gamma$ score used in the reconstruction of $\gamma$ candidates.
A benefit of applying semantic segmentation in such a way, instead of, for example, an a more complicated likelihood algorithm during the reconstruction, is that the reconstruction and associated uncertainties can be compared at every point in the reconstruction. 

\subsection{Hit selection}
\label{sec:hitselection}
Clusters are a set of nearby hits in the same plane; those not used to create tracks (generally charged pions, protons, and muons) are made available for the electromagnetic shower reconstruction.
The hits in the cluster must have an associated time within 25 ns of the identified muon track.
Additionally, low-activity clusters that have less than three photo-electrons per hit are removed, as they are most likely to originate from cross-talk.
Moreover, clusters in the HCAL are not included in the available hits.

For the semantic segmentation based reconstruction, we place a requirement that the considered hits must have probability to be electromagnetic-like, as determined by the MLA, greater than a threshold of $0.5$ which is the expected optimum due to the interpretation of the MLA output as a probability of the given class.
This improves the purity of the sample during the formation of available clusters.
This additional filter in the hit selection is the only difference between the baseline reconstruction and the reconstruction that uses ML.

\subsection{Electromagnetic shower (EM) candidate reconstruction}
\label{sec:emshowerreconstruction}

\begin{figure}[!htb]
\centering
\begin{subfigure}{0.5\textwidth}
  \centering
  \includegraphics[height=4.5cm, width = 6.8cm]{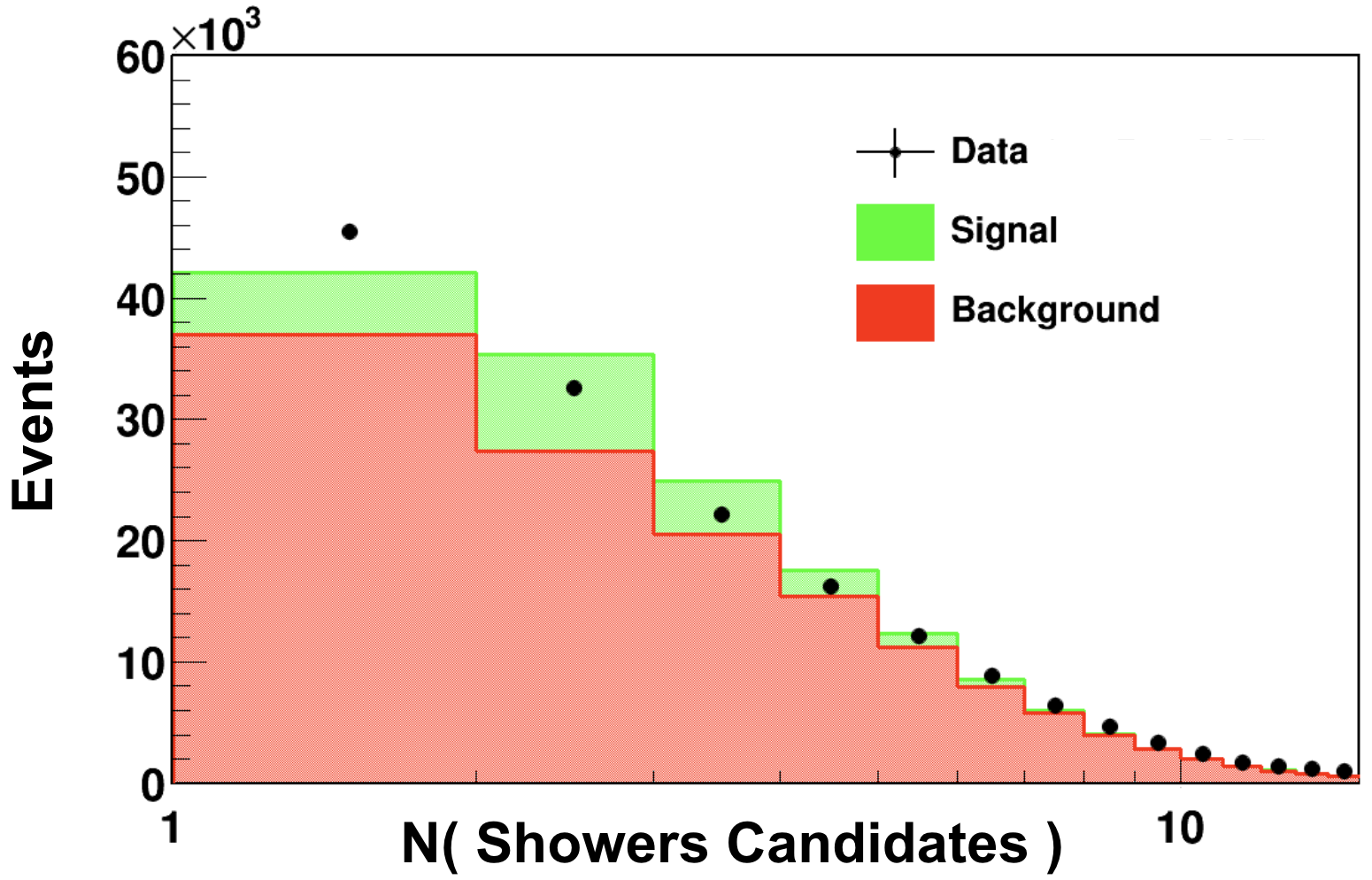}
  \label{fig:ShowerCandidates_NoML}
  \end{subfigure}%
\begin{subfigure}{0.5\textwidth}
  \centering
  \includegraphics[height=4.5cm, width = 6.8cm]{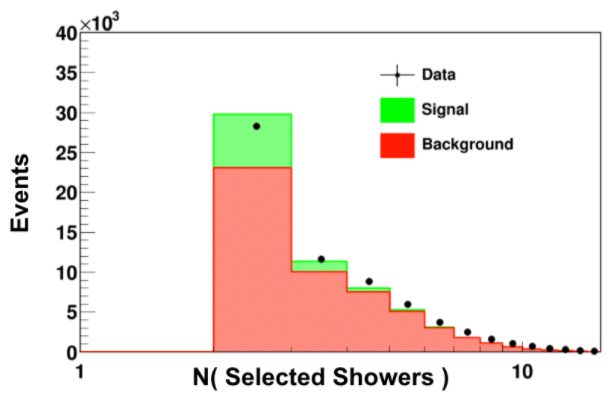}
  \label{fig:SelectedShowers_NoML}
\end{subfigure}%
\caption{Baseline reconstruction in the semi-exclusive analysis: a) Number of electromagnetic shower candidates selected by a Bayesian algorithm. b) Showers from a) for which a three-dimension reconstruction is possible.  The number of events in the simulation is normalized to the data. Systematic uncertainties are not shown. 
}
\label{fig:Showers_GoodBlobs_NoML}
\end{figure}

Clusters are then assigned to electromagnetic showers via a Bayesian algorithm that groups the available clusters of the X-view into conical regions emanating from the neutrino interaction vertex.
The assumption is that these clusters arise from an electromagnetic shower originating at the neutrino interaction vertex.
In Fig.\ref{fig:Showers_GoodBlobs_NoML}a we show the results of the Bayesian algorithm applied to the semi-exclusive analysis.
This allows one to associate a group of clusters with a shower that is characterized by the angle between the shower and the longitudinal axis of the detector.
These groupings of clusters are then considered in the next stage of the electromagnetic shower reconstruction.

Clusters of the U-view and V-view are then included to form electromagnetic shower candidates.
The position of the centroid for the electromagnetic shower candidate is given by $\vec{R}=\frac{\sum_c E_c \vec{r}_c}{\sum_c E_c}$ where $E_c$ is the energy of the cluster in photo-electrons and $\vec{r}_c$ is the position of the vector with respect to the interaction vertex.

\begin{figure}[!htb]
\centering
\begin{subfigure}{0.5\textwidth}
  \centering
  \includegraphics[height=4.5cm, width = 6.8cm]{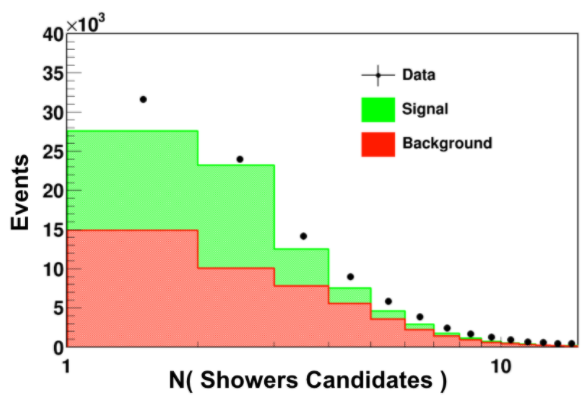}
  \label{fig:GoodBlobs}
  \end{subfigure}%
\begin{subfigure}{0.5\textwidth}
  \centering
  \includegraphics[height=4.5cm, width = 6.8cm]{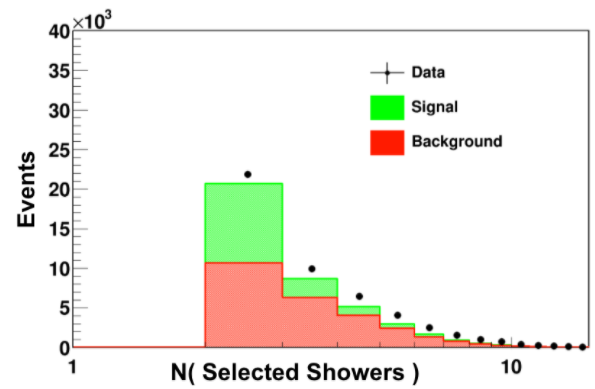}
  \label{fig:SelectedShowers}
\end{subfigure}%
\caption{Machine learning reconstruction in the semi-exclusive analysis: a) Number of electromagnetic shower candidates selected by a Bayesian algorithm. b) Showers from a) for which a three-dimension reconstruction is possible.  The number of events in the simulation is normalized to the data. Systematic uncertainties are not shown. Shown are the distributions before tuning of the Monte-Carlo simulation and determination of the systematic uncertainty.}
\label{fig:Showers_GoodBlobs}
\end{figure}

The first topological condition for the electromagnetic shower candidate is that each candidate must have hits in at least two views.
This allows three-dimensional reconstruction of the direction of the shower.
In Fig.\ref{fig:Showers_GoodBlobs_NoML}b these are referred to as "Selected Showers".
The corresponding figures for the electromagnetic shower candidate reconstruction with ML are in Fig.\ref{fig:Showers_GoodBlobs} demonstrating improved purity.

\subsection{$\gamma$ reconstruction}
\label{sec:gammareconstruction}

The electromagnetic shower candidates are then reconstructed into $\gamma$ candidates.
As part of this reconstruction, a second topological condition is applied.
This requirement, that the distance from the interaction vertex to the centroid of the shower candidate be at least $14$ cm, is motivated by the conversion length of a $\gamma$ in scintillator being $40$ cm.

In analyses of the NuMI Low Energy Beam dataset \cite{Altinok:2017xua}\cite{Aliaga:2015wva}\cite{Coplowe:2020yea} the second requirement was only applied to the highest energy $\gamma$ candidate.
In the NuMI Medium Energy Beam dataset, neutral pions are produced at higher energy, as are the other particles produced in the interaction. 
This results in complications in the reconstruction of $\gamma$s due to the production of more particles in the neutrino interaction, increased interactions of energetic particles with the detector material, and the production of high energy $\gamma$s through stochastic processes in the electromagnetic shower.

These complications, exacerbated by passive material in the detector, can result in the production of a high energy $\gamma$ early in the electromagnetic shower, the production of more neutrons and the production of energetic particles not originating from the neutrino vertex.
This produces more $\gamma$ candidates during the electromagnetic shower reconstruction and increases the energy of the candidate $\gamma$s, both due to $\gamma$s being at higher energy and due to overlap between the electromagnetic shower from the $\gamma$ and other energy from non-electromagnetic sources.
Additionally, the more energetic neutral pions often have a narrower opening angle between the $\gamma$s.
The strategy that we have followed to address such challenges for neutral pion reconstruction in the NuMI Medium Energy Beam dataset is to improve the selection of $\gamma$ candidates during the $\gamma$ reconstruction as described below.

\begin{figure}[!htb]
\centering
\begin{tabular}{cccc}
  \includegraphics[width=0.3\textwidth]{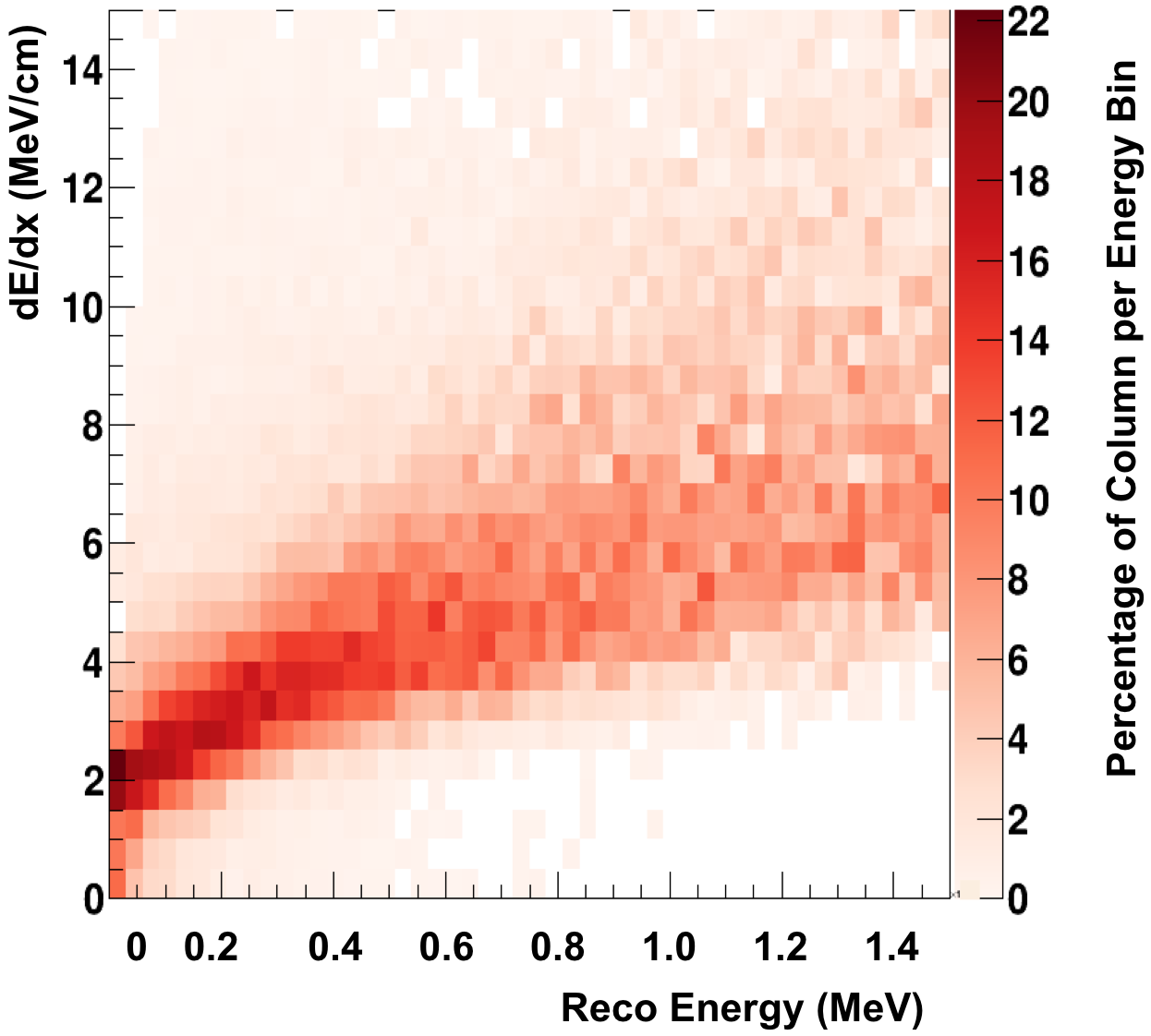} &
  \includegraphics[width=0.3\textwidth]{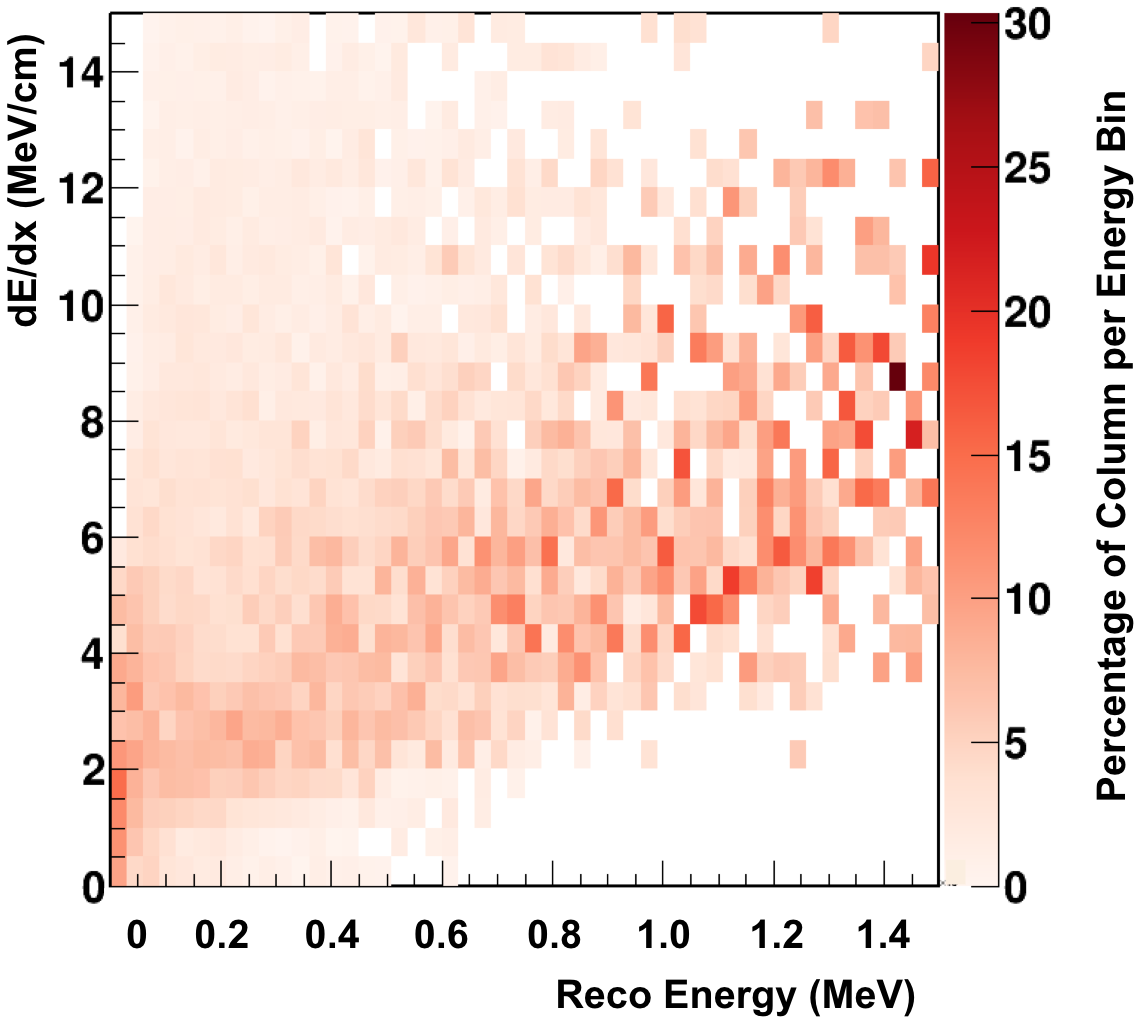} &
  \includegraphics[width=0.3\textwidth]{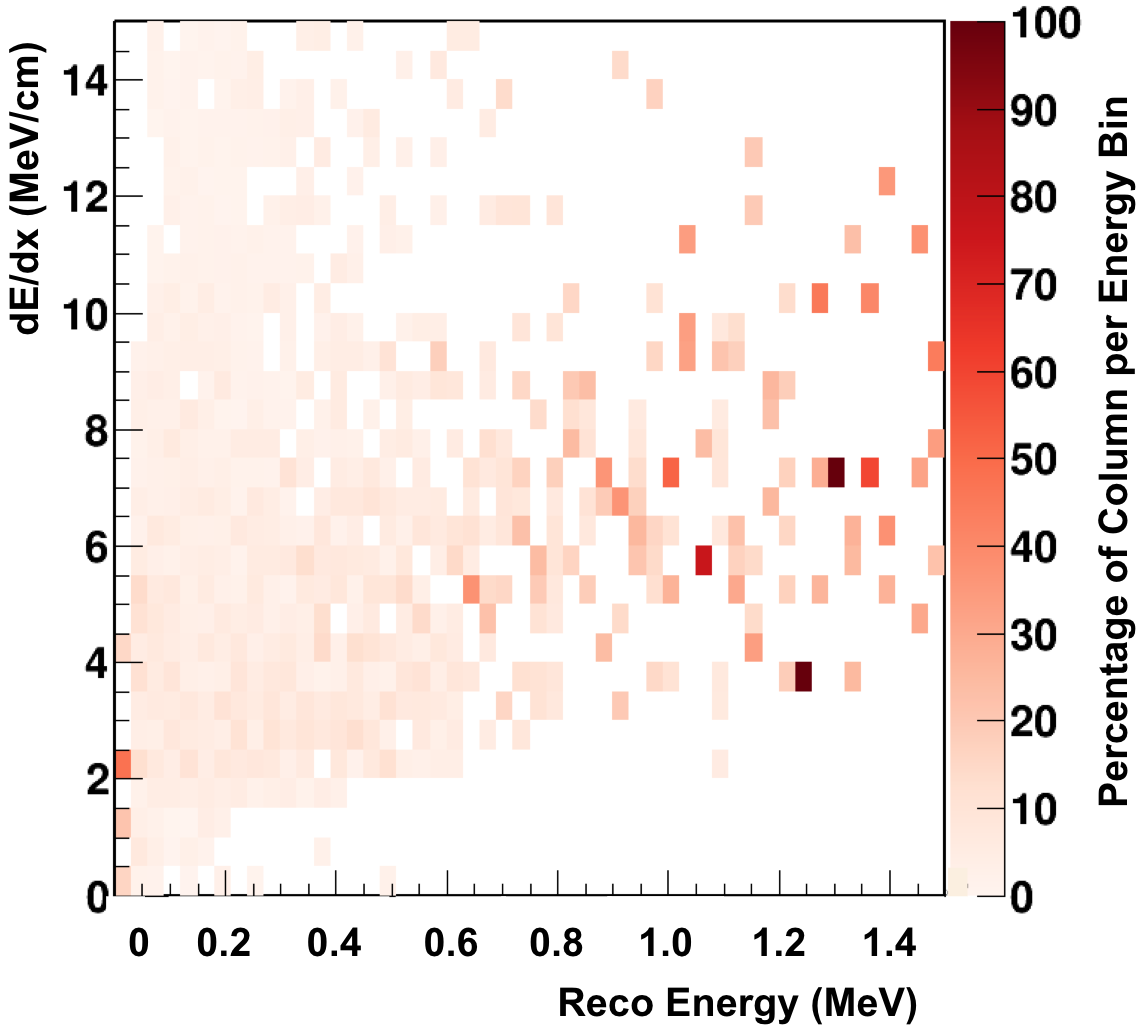} \\
(a) $\gamma$ & (b) Pion (+)  & (c) Pion (-)   \\[6pt]
\end{tabular}
\begin{tabular}{cccc}
  \includegraphics[width=0.3\textwidth]{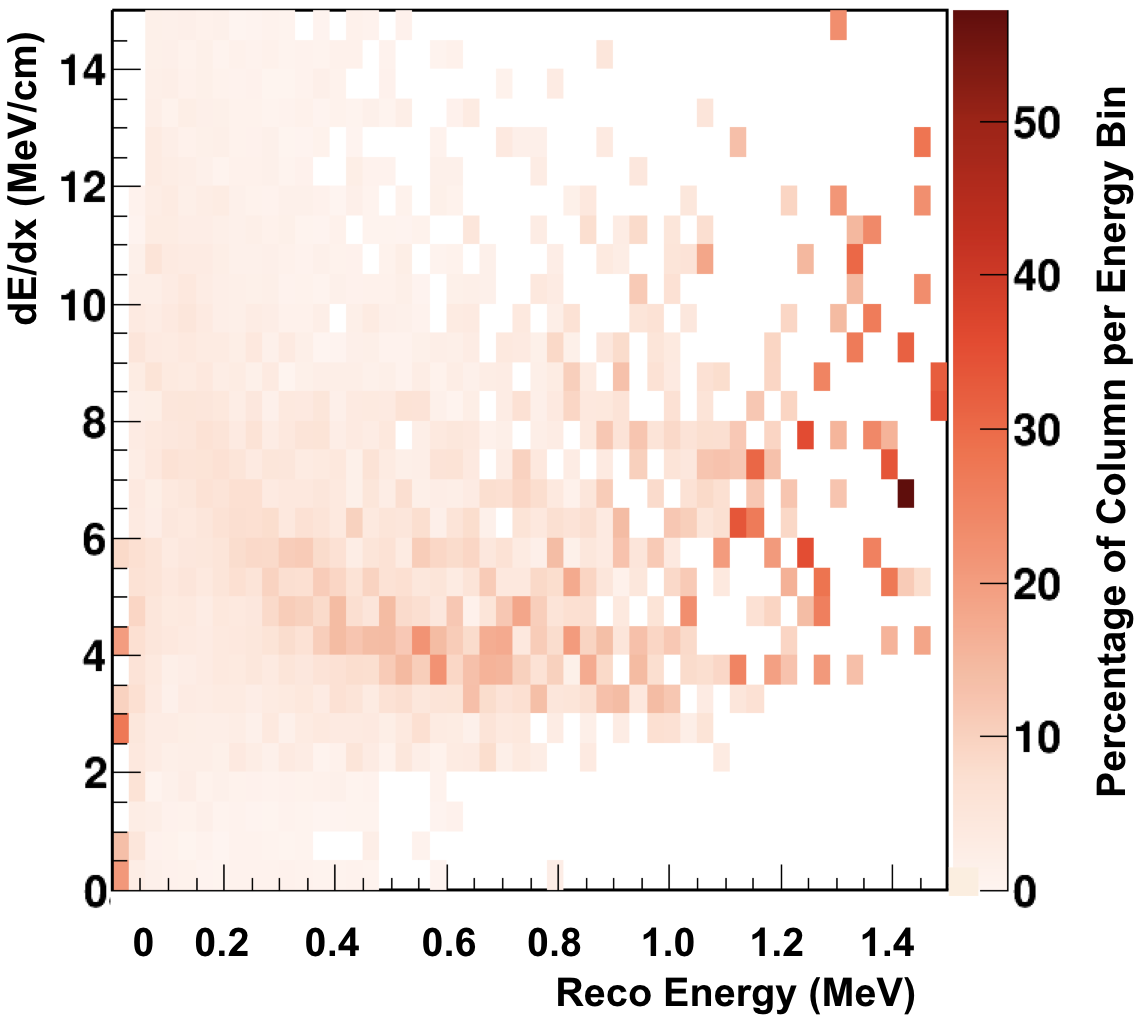} &
  \includegraphics[width=0.3\textwidth]{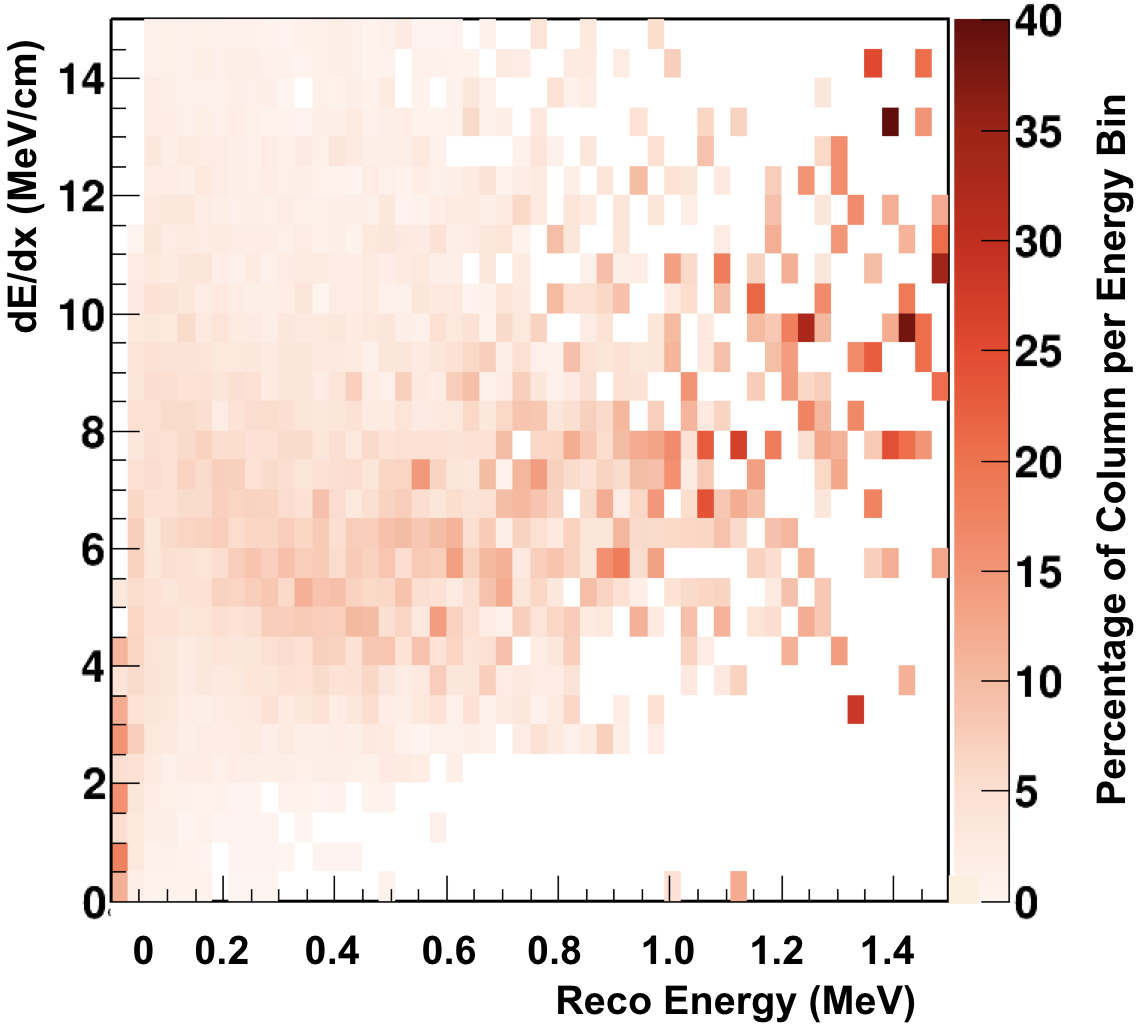}\\
  (d) Proton  & (e) Neutron  \\[6pt]
\end{tabular}
\caption{Reconstructed energy versus $\frac{dE}{dx}$ for the primary $\gamma$ candidate in simulated events in the semi-exclusive analysis with {\bf the baseline reconstruction}.   The candidates are classified according to the particle which produced the electromagnetic shower, (a) to (e). There is a clear correlation between energy and $\frac{dE}{dx}$ for candidates produced by true photons (a).  }
\label{fig:particle_RecoEnergy_dedx_noML}
\end{figure}

\begin{figure}[!htb]
\centering
\begin{tabular}{cccc}
  \includegraphics[width=0.3\textwidth]{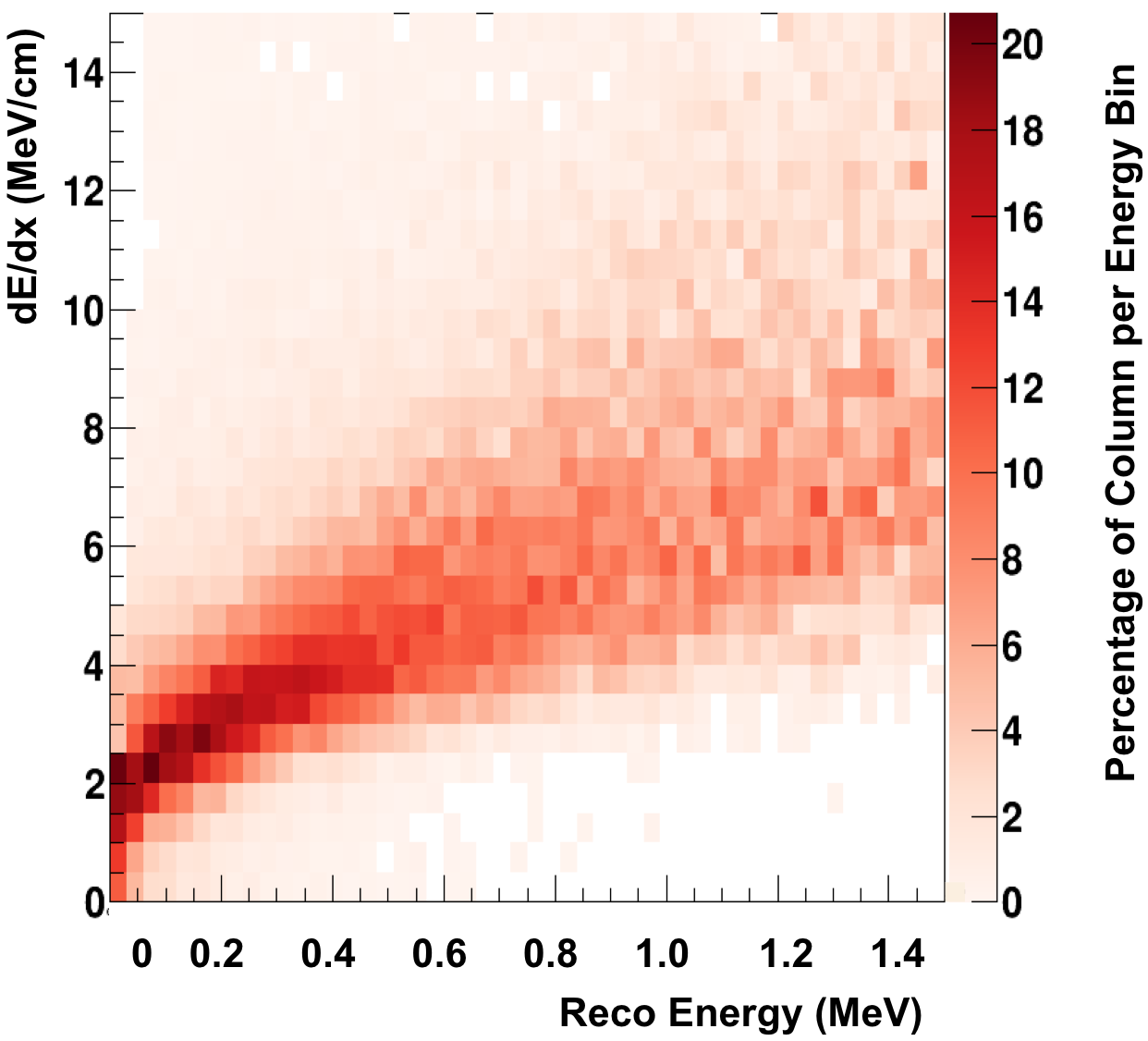} &
  \includegraphics[width=0.3\textwidth]{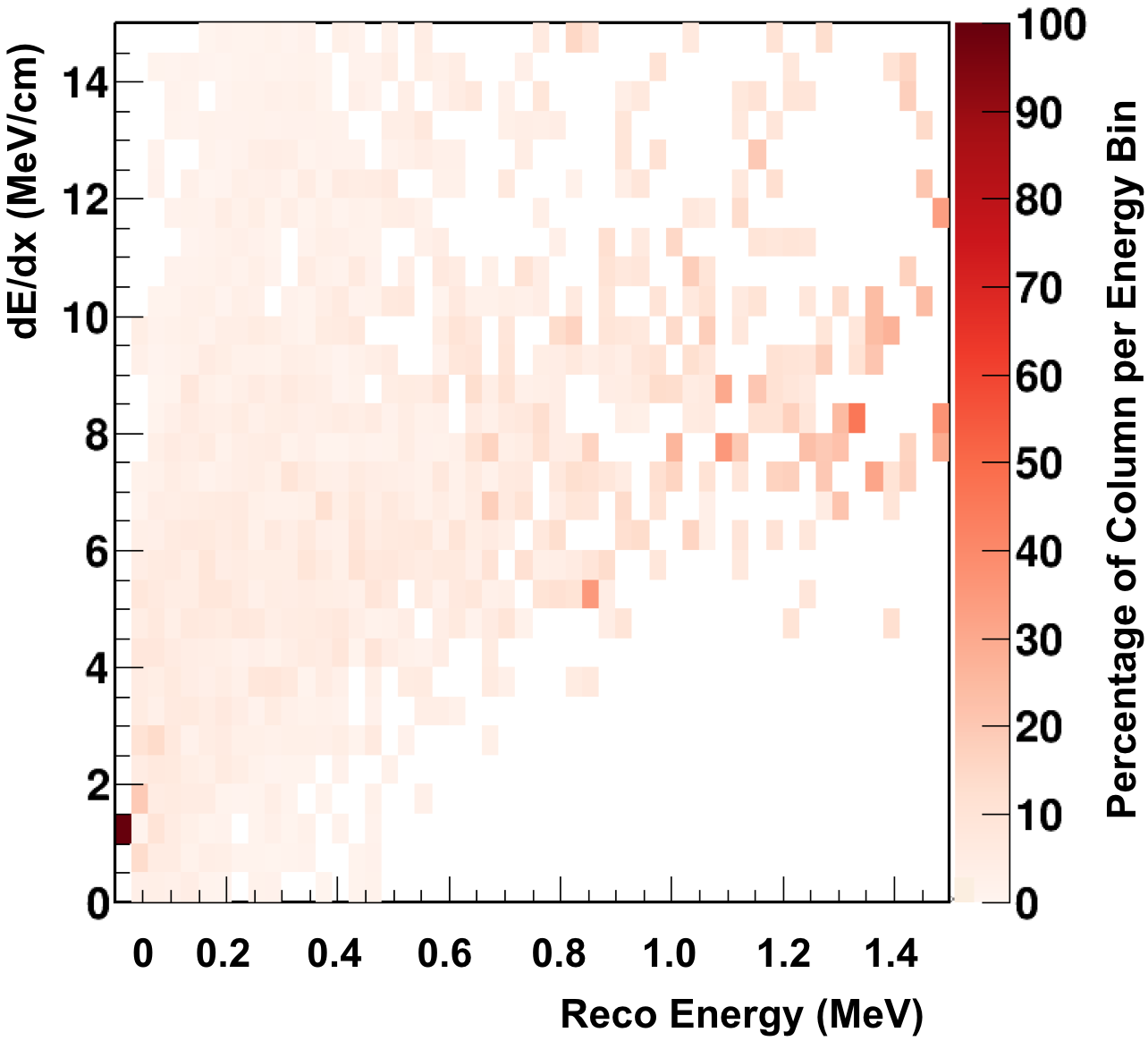} &
  \includegraphics[width=0.3\textwidth]{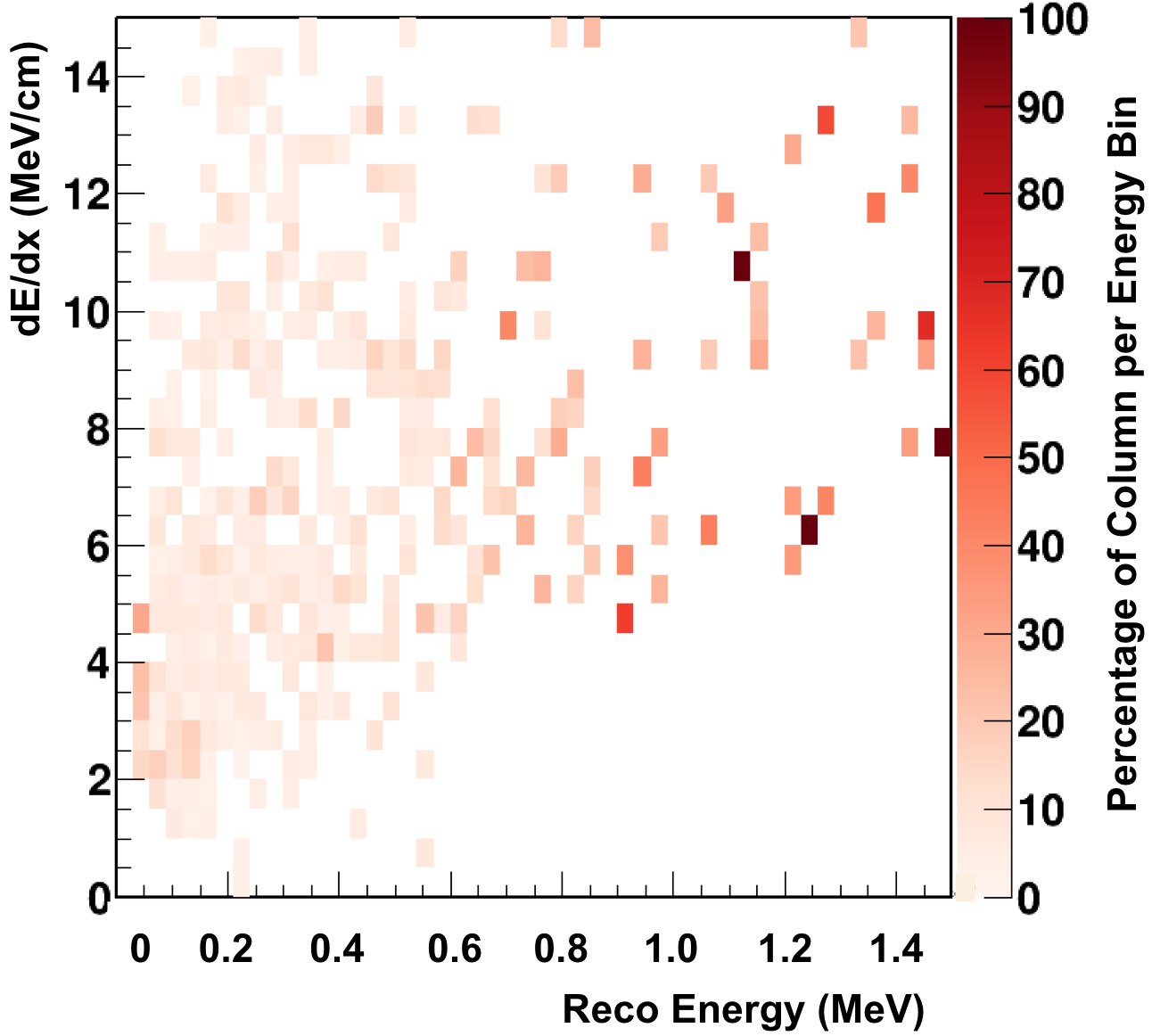} \\
(a) $\gamma$  & (b) Pion (+) & (c) Pion (-)  \\[6pt]
\end{tabular}
\begin{tabular}{cccc}
  \includegraphics[width=0.3\textwidth]{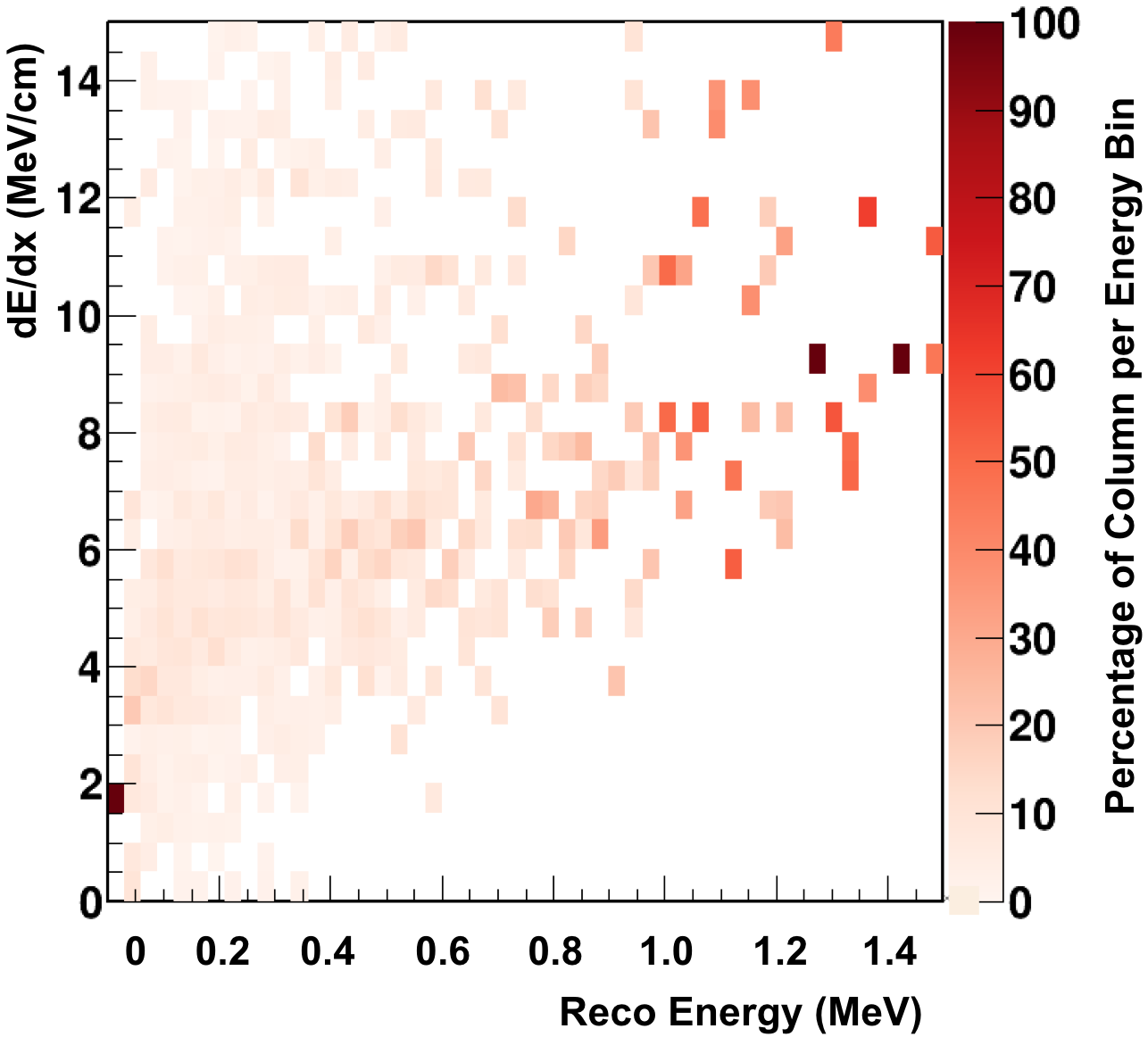} &
  \includegraphics[width=0.3\textwidth]{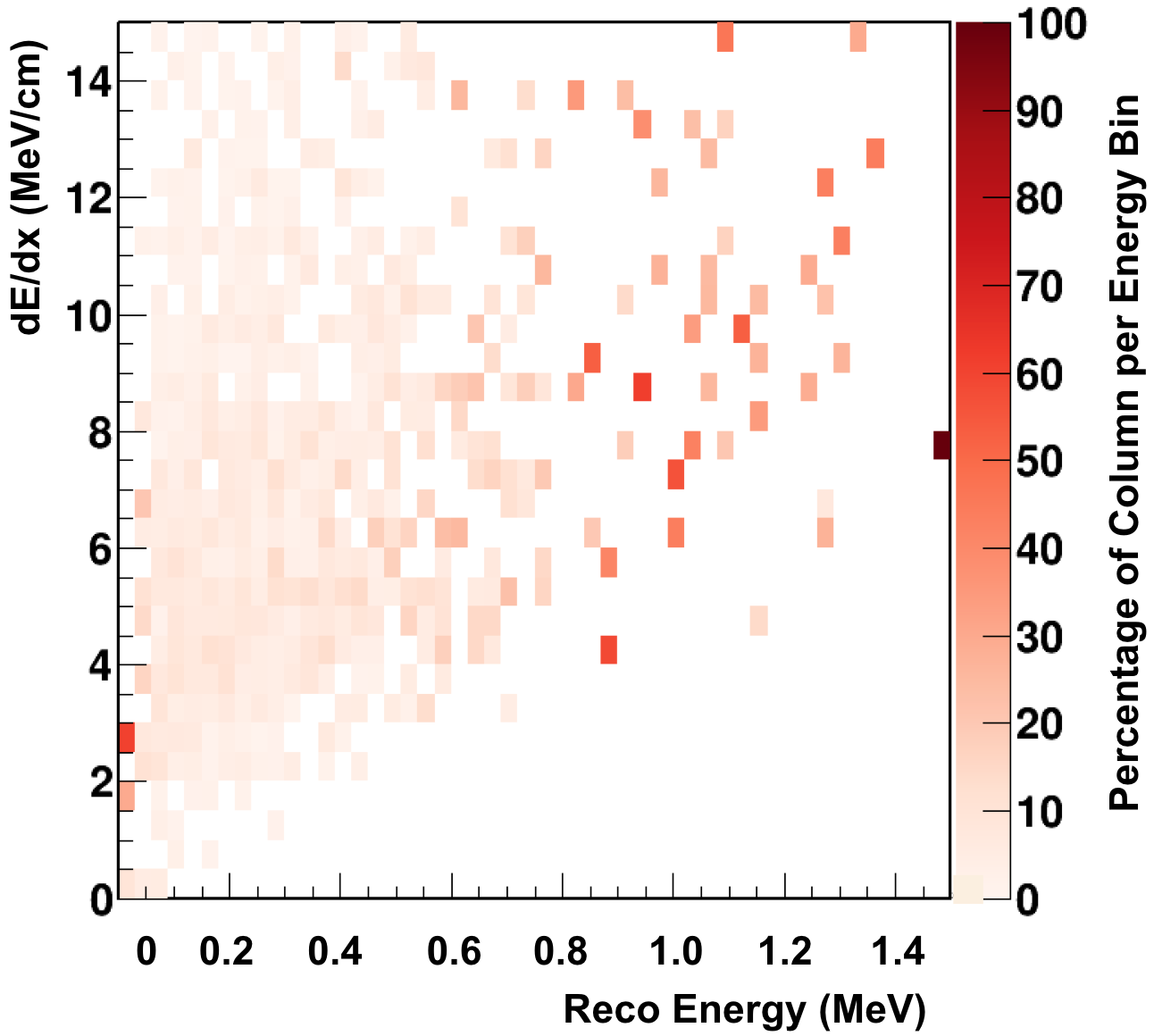}\\
  (d) Proton  & (e) Neutron  \\[6pt]
\end{tabular}
\caption{Reconstructed energy versus $\frac{dE}{dX}$ for the primary $\gamma$ candidate in simulated events in the semi-exclusive analysis with {\bf the ML reconstruction}.   The candidates are classified according to the particle which produced the electromagnetic shower, (a) to (e). There is a clear correlation between energy and $\frac{dE}{dX}$ for candidates produced by true photons (a).  }
\label{fig:particle_RecoEnergy_dedx_ML}
\end{figure}

\begin{figure}[!htb]
\centering
\begin{tabular}{cccc}
  \includegraphics[height=7.0cm, width = 7cm]{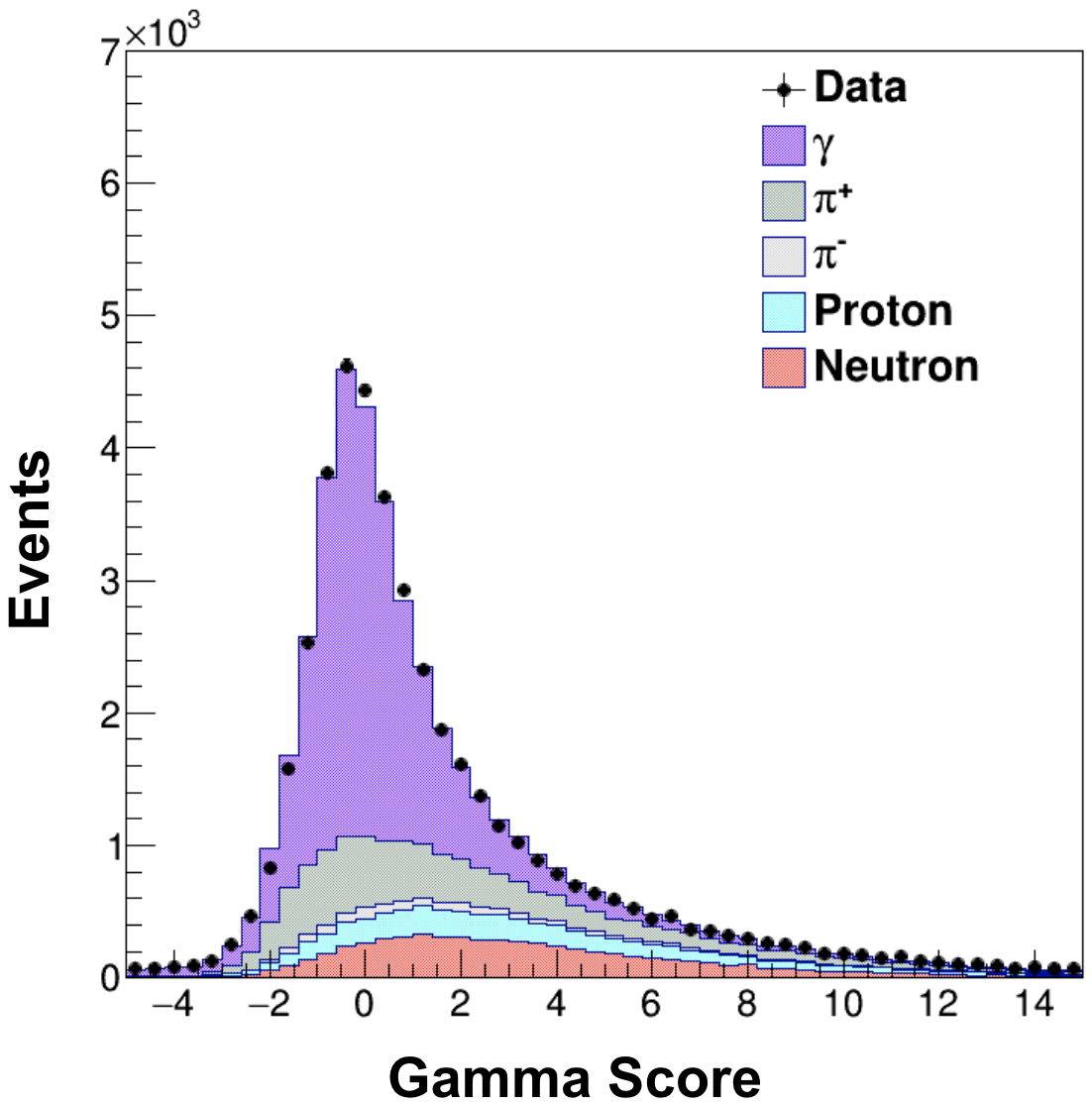} &
  \includegraphics[height=7.0cm, width = 7cm]{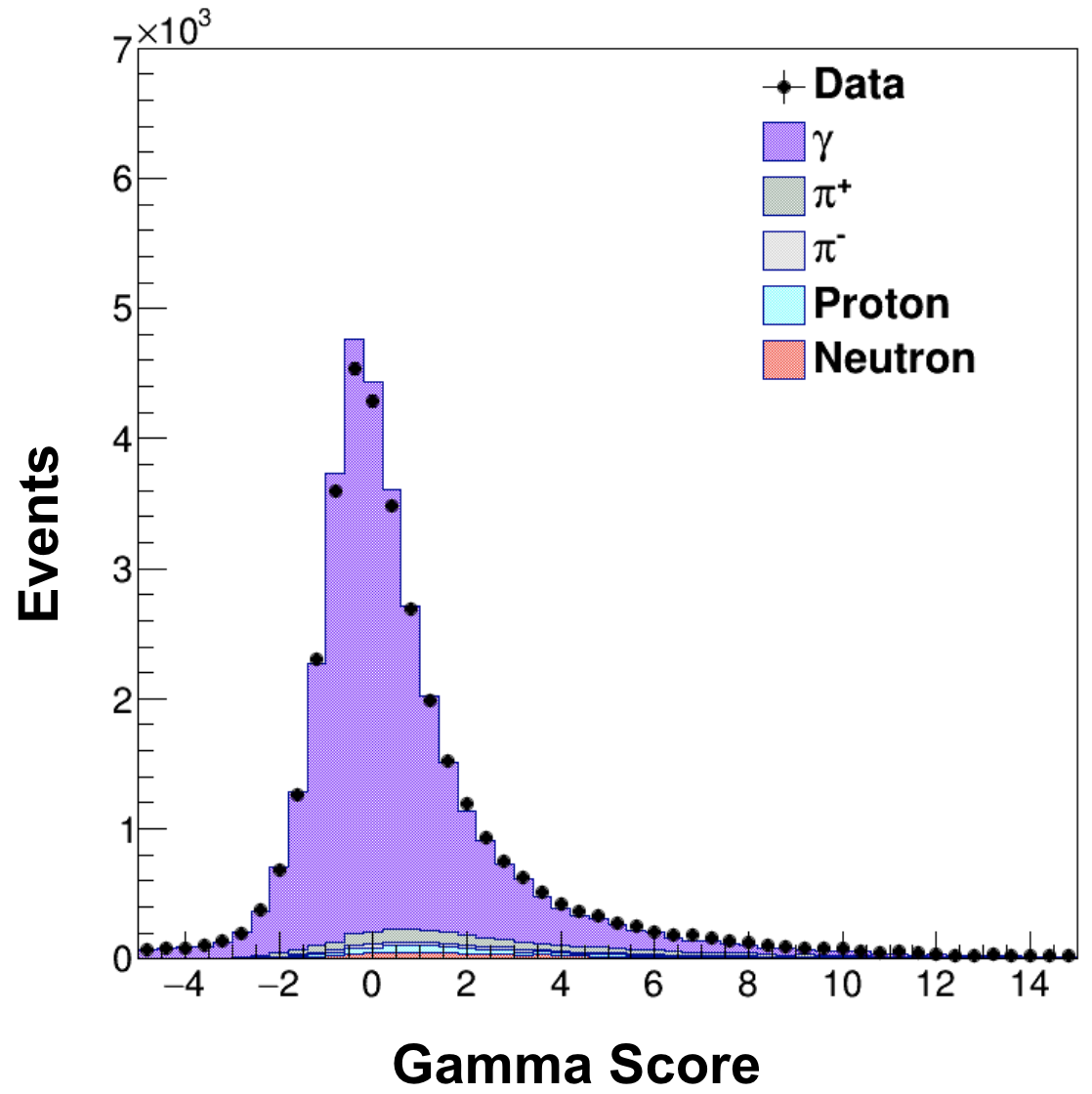} \\
 (a) Baseline reconstruction   & (b) Machine learning reconstruction \\[6pt]
\end{tabular}
\begin{tabular}{cccc}
  \includegraphics[height=7.0cm, width = 7cm]{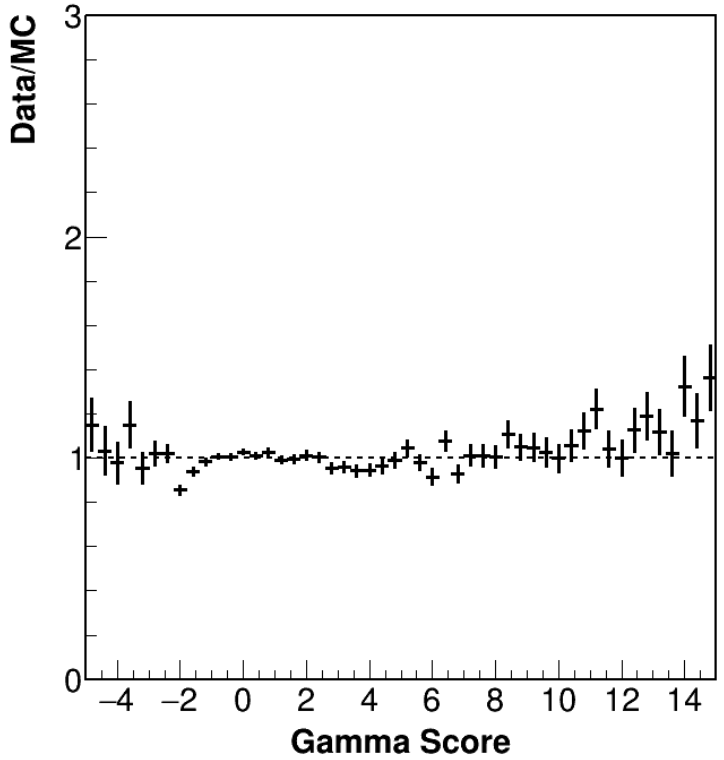} &
  \includegraphics[height=7.0cm, width = 7cm]{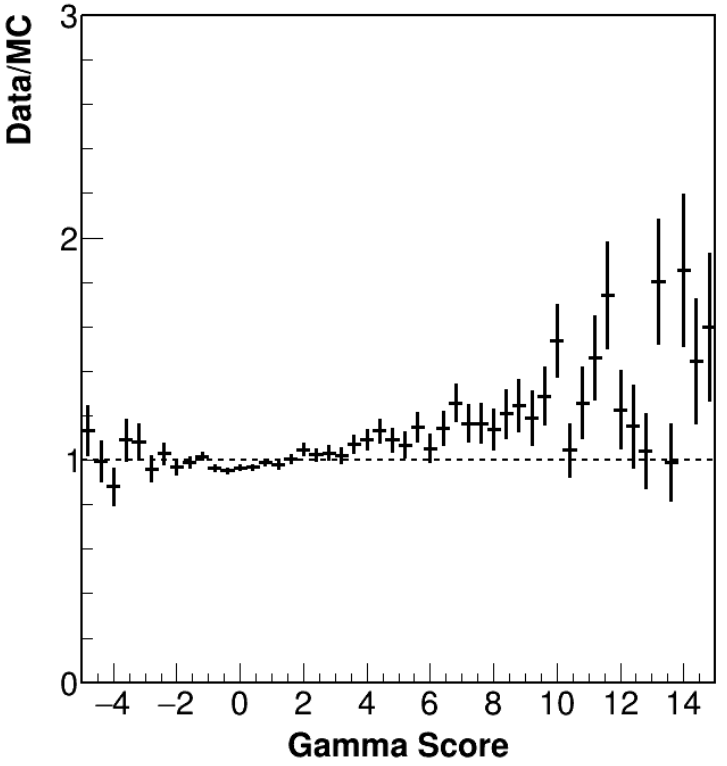}\\
  (c) Ratio between data and simulation for  & (d) Ratio between data and simulation \\ the baseline reconstruction  & for the ML reconstruction \\[6pt]
\end{tabular}
\caption{$\Gamma_{score}$ for the semi-exclusive analysis in data and simulation.  The simulation is classified based on the true particle that contributed the greatest fraction of energy to the $\gamma$ candidate.
In the baseline reconstruction a $\Gamma$ score of less than $2$ is a signature of a photon.
When using the ML prediction, the $\Gamma$ score is not a useful metric to improve the purity of the $\gamma$ reconstruction.
The number of events in the simulation is normalized to the data.
}
\label{fig:photon_score}
\end{figure}

A third topological feature was developed for the $\gamma$ candidates.
This feature is based on the total deposited visible energy in the $\gamma$ candidate per unit length, given by:
\begin{equation}
\frac{dE}{dx} = \frac{E_{vis}}{l}
\end{equation}
where $E_{vis}$ is the visible energy and $l$ is the length of the $\gamma$ candidate.
Figs.~\ref{fig:particle_RecoEnergy_dedx_noML} and~\ref{fig:particle_RecoEnergy_dedx_ML} show reconstructed energy versus $\frac{dE}{dx}$ for $\gamma$ candidates using the baseline and ML-based reconstructions, respectively.  
In both cases, $\gamma$ candidates created by true photons have a clear correlation between  reconstructed energy and $\frac{dE}{dx}$.  
The mean and standard deviation of the $\frac{dE}{dx}$ distributions for $\gamma$ candidates produced by true photons are used to create a $\Gamma$ score given by:
\begin{equation}
    \Gamma_{score}=\frac{\frac{dE}{dx}_{obs} - \langle \frac{dE}{dx} \rangle_i}{\sigma_{\frac{dE}{dx},i}},
\end{equation}
where $\frac{dE}{dx}_{obs}$ is the $\frac{dE}{dX}$ of a $\gamma$ candidate and $\langle \frac{dE}{dx}\rangle_i$ and  $\sigma_{\frac{dE}{dx},i}$ are the mean and standard deviation of the simulated dE/dx distribution for true photons, evaluated in the reconstructed energy bin corresponding to the $\gamma$ candidate.  The distribution of $\gamma$ candidate scores in the semi-exclusive analysis is shown in Fig. \ref{fig:photon_score} for both the baseline and ML reconstruction.  

For $\gamma$ candidates that are not associated with photons, the $\Gamma$s score is observed to be very asymmetric, while the score for true $\gamma$s behaves approximately as a Gaussian. In the baseline reconstruction, this provides the third topological requirement that the $\gamma$ candidate has a score of $\Gamma_{ score}<2$. Having a score of $\Gamma_{score}>2$ indicates that the $\gamma$ candidate is not likely to be a photon.  For the ML-based reconstruction, $\Gamma$ score is not a useful discriminator, presumably because $\frac{dE}{dx}$ has already been taken into account by the ML algorithm.

In Fig. \ref{fig:photon_score} we see that the match between simulation and data is similar for both techniques for $\gamma$ candidates with $\Gamma$ score $\leq$ 2.0, but that data is more often under simulation for the baseline reconstruction and over simulation for the semantic segmentation based reconstruction in the peak area of -2 to 2.
The neutral pion reconstruction includes the cut on $\Gamma$ score of less than $2$ and this is reflected in the final plots and numbers.
In the high tail, with high uncertainty we see that, the semantic segmentation based reconstruction over predicts the data with a linear dependence on the $\Gamma$ score.

Finally, the energy of the $\gamma$ candidates is corrected by a scale factor to correct the simulation for the detector response due to electromagnetic activity.
To do this we used energetic $\gamma$s from the semi-inclusive neutral pion analysis; as described in \cite{Valencia:2019mkf}.
Analysis of this scale factor is an important contribution to the systematic uncertainty of analyses which depend on measurements of electromagnetic activity.

\subsection{Neutral pion reconstruction}
\label{sec:pi0reconstruction}

\begin{figure}[!htb]
\centering
\begin{tabular}{cccc}
  \includegraphics[height=7.5cm, width = 7.5cm]{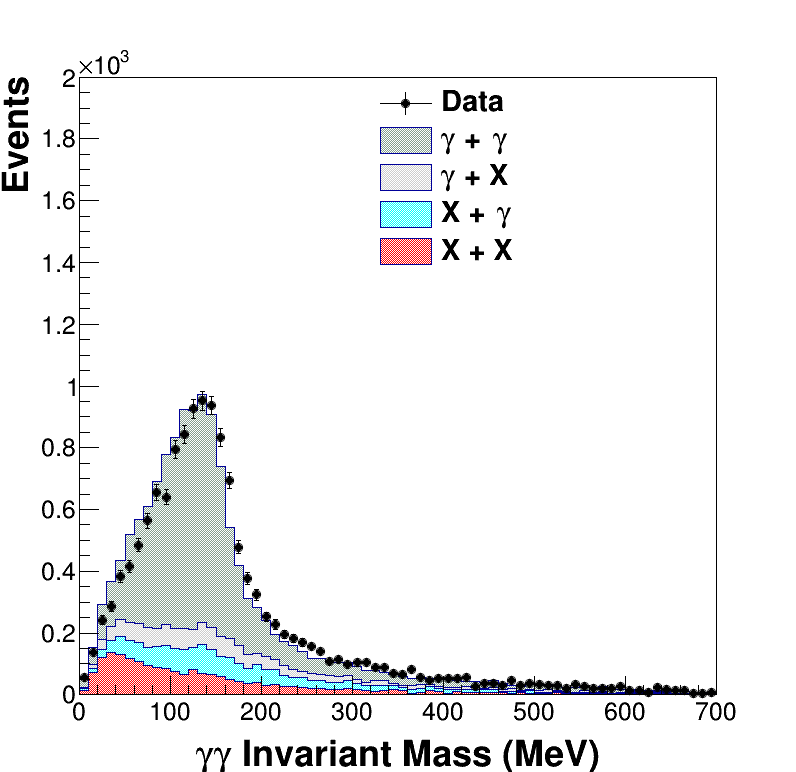} &
  \includegraphics[height=7.5cm, width = 7.5cm]{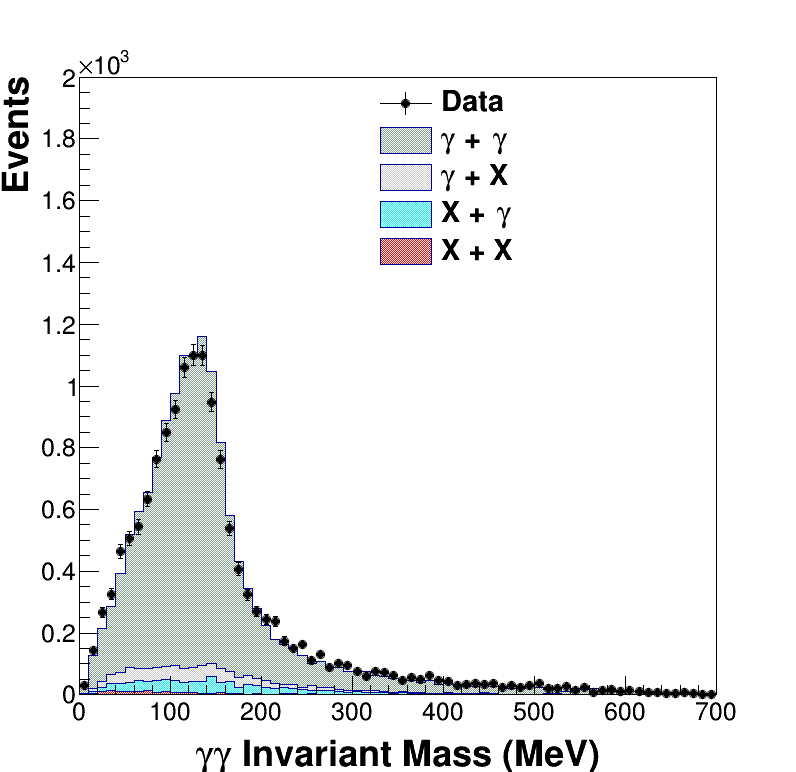} \\
 (a) Baseline reconstruction   & (b) Machine learning reconstruction \\[6pt]
\end{tabular}
\begin{tabular}{cccc}
  \includegraphics[height=6.5cm, width = 7.0cm]{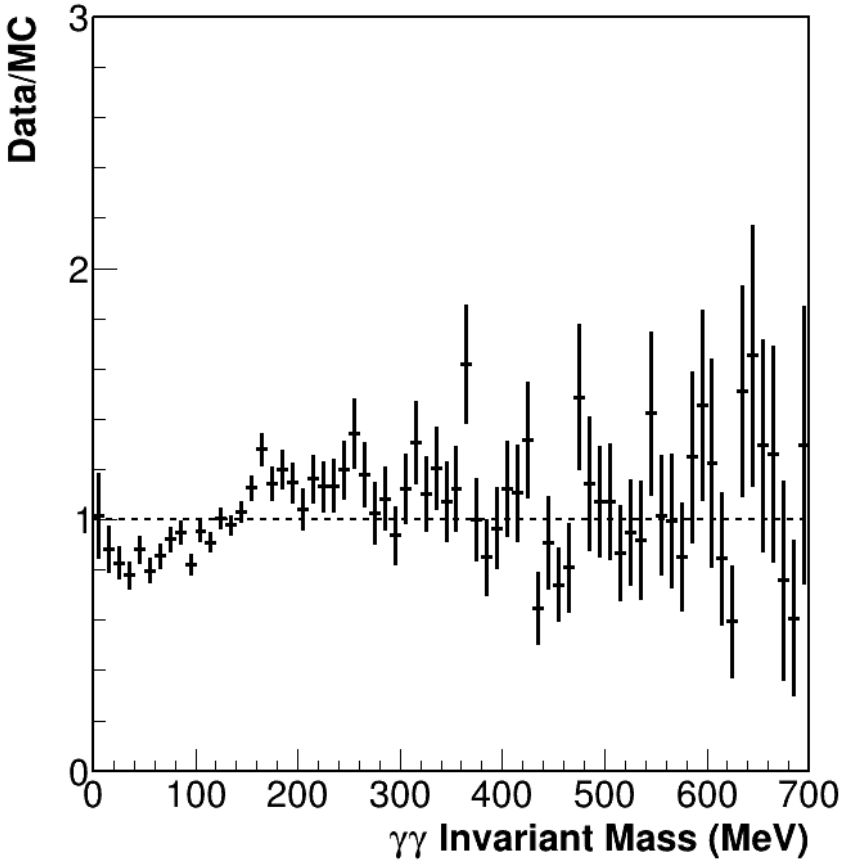} &
  \includegraphics[height=6.5cm, width = 7.0cm]{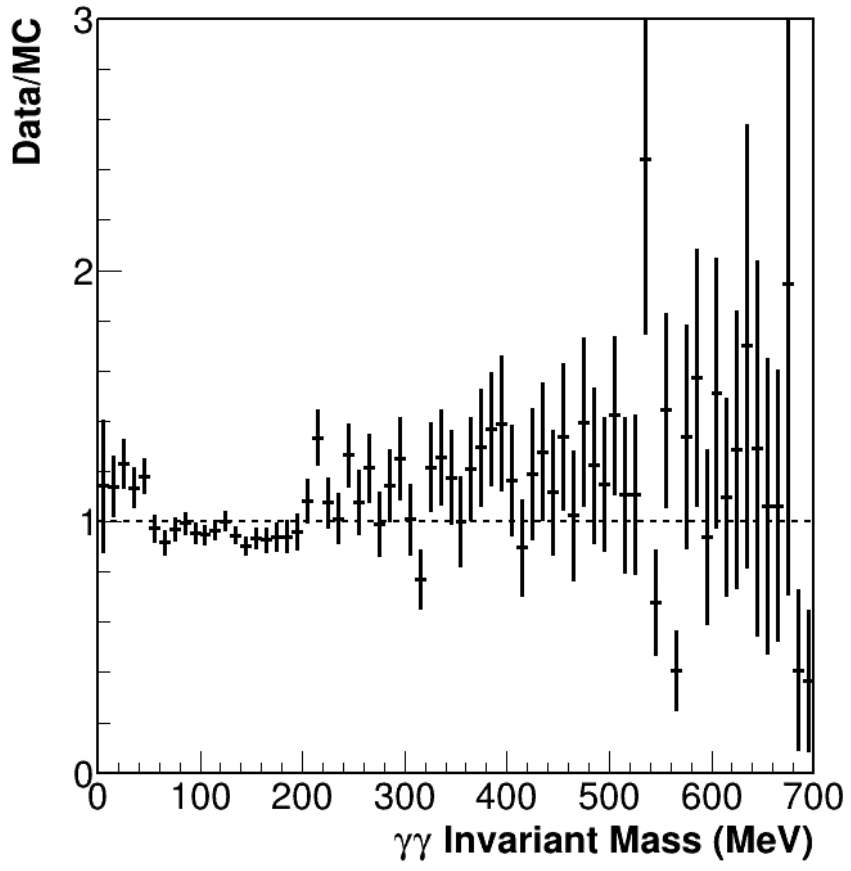}\\
  (c) Ratio between data and simulation for  & (d) Ratio between data and simulation \\ the baseline reconstruction  & for the ML reconstruction \\[6pt]
\end{tabular}
\caption{Neutral pion invariant mass distribution between $[0-700]$ MeV for the semi-exclusive analysis in data and simulation. The simulation is classified based on the true particle that contributed the greatest amount of energy to the primary and secondary $\gamma$ candidates. $X$ corresponds to the absence of a $\gamma$.  Note the significant improvement in the size of the $\gamma+\gamma$ category and decrease in all other categories when semantic segmentation provides context. From this result our final event selection are events between $[60-200]$ MeV as shown on Table\ref{table_0_700_60_200}. The number of events in the simulation is normalized to equal that of the data. }
\label{fig:blobsMass_BarbID}
\end{figure}

The neutral pion analyses select events with one and only one neutral pion, so the first part of the neutral pion reconstruction is to require that two and only two reconstructed $\gamma$s remain after $\gamma$ reconstruction.  The second component of the neutral pion reconstruction is the calculation of the neutral pion candidate invariant mass, $m_{\gamma\gamma}$, and selection based on that invariant mass.
The neutral pion candidate invariant mass is given by
\begin{equation}\label{InvMass}
    m_{\gamma\gamma} = \sqrt{2 E_{\gamma 1}E_{\gamma 2} \left( 1 - \cos \Theta_{\gamma\gamma}\right)}
\end{equation}
where $E_{\gamma 1(2)}$ is the reconstructed energy of the most (least) energetic $\gamma$ and $\Theta_{\gamma\gamma}$ is the opening angle between the reconstructed $\gamma$s.
The neutral pion invariant mass is shown in Fig.~\ref{fig:blobsMass_BarbID} for the semi-exclusive analysis, where we can see that the two $\gamma$s are most often reconstructed into a neutral pion candidate with $60<m_{\gamma\gamma}<200$ MeV.
There is also an evident shift between data and simulation whose source is currently unknown, but it is present with both the baseline and ML reconstructed samples, and is also present in the semi-inclusive analysis.
That analysis has a different background than the semi-inclusive, due to both the change of signal definition and the change in event selections and is discussed in Appendix~\ref{sec:semiinclusive}.  

The simulated invariant mass distribution in Figure~\ref{fig:blobsMass_BarbID} is divided according to the true identity of the particles creating the photon candidates.  The percentage of events in each category is given in Table~\ref{table_0_700_60_200} for the entire 0-700 MeV mass range and 
for the 60-200 MeV range where most true $\pi^0$ candidates reside.  Within the 60-200 MeV range, $70.7\pm0.9$\% of reconstructed $\pi^0$ candidates are reconstructed from two true photons in the simulation, while this number rises to $89.3\pm0.7$\% with the ML-based algorithm.

While the neutral pion reconstruction obviously also provides the neutral pion kinetic energy, momentum and angle with respect to the neutrino beam axis, this paper will not include a detailed discussion of such quantities other than to note the evident relationship between improving the reconstruction of the $\gamma$s and improving the measurement of these kinematic quantities, which will be the subject of future MINERvA publications.  

\begin{table}[!ht]
\centering
\begin{tabular}{l|cc|cc}
 \toprule
\textbf{Event}    & \textbf{Full Distribution} &  \textbf{$[0-700]$ MeV}  & \textbf{Selection Distribution} &   \textbf{$[60-200]$ MeV} \\
  \textbf{Category}                & Baseline                   & Machine Learning         & Baseline                        & Machine Learning\\             
 \midrule
 $\gamma + \gamma$             & $62.5\pm0.8$      & $87.1\pm0.6$    & $70.7\pm0.9$     & $89.3\pm0.7$ \\            
 $\gamma$ + X                  & $11\pm2$      & $6\pm2$     & $9\pm2$      & $5\pm3$\\
 X + $\gamma$                  & $12\pm2$      & $6\pm3$     & $10\pm2$     & $5\pm3$\\
 X + X                         & $14\pm2$      & $1^{+6}_{-1}$     & $10\pm2$     & $1^{+7}_{-1}$ \\
 \midrule
 Total                         & 100       & 100     & 100      & 100 \\ 
\bottomrule
\end{tabular} 
  \caption
  {Fraction (in per cent) of simulated events by category in the neutral pion invariant mass distributions between $[0-700]$ MeV and between $[60-200]$ MeV which is the region where the efficiency of the selection is better, from Fig. \ref{fig:blobsMass_BarbID} . uncertainties are statistical. }
\label{table_0_700_60_200}
\end{table}

%% file: sec_conclusion.tex
\FloatBarrier 
\section{Discussion \& Conclusions}
\label{sec:conclusions}
 
 The new approach keeps more real neutral pion signal and rejects more background than the baseline approach.
 This is clearly observed in Figs.~\ref{fig:photon_score} and \ref{fig:blobsMass_BarbID} (with associated Table~\ref{table_0_700_60_200}).
 The reconstruction purity is observed to increase from $70.7\pm0.9$\% to $89.3\pm0.7$\% in the signal region [60-200] MeV of Fig. \ref{fig:blobsMass_BarbID} as shown in Table \ref{table_0_700_60_200}, while the efficiency of the reconstruction increases by approximately 40\%.
 The only change in the new approach is the application of semantic segmentation to provide context in the hit filter stage of the reconstruction.
 This improvement results in improvements to both the efficiency and purity in the charged current neutral pion production analysis, which is the subject of a forthcoming publication by MINERvA.  
 

 \subsection{Discussion}
 \label{sec:discussion}

A similar technique to the one described in Section \ref{sec:pi0_rec_ML} was used by \uboone, both for semantic segmentation of electromagnetic showers\cite{Adams:2018bvi} and for the application of that semantic segmentation as a hit filter for a neutral pion reconstruction\cite{Adams:2019law}.
While the techniques are similar in that they are both based on semantic segmentation, the experimental environments are quite different.  The mean energy of NuMI Medium Energy Beam dataset for \minerva is $\langle E_\nu \rangle \sim 6$ GeV while for \uboone $\langle E_\nu \rangle \sim 800$ MeV, causing \minerva to have a much richer sample in the resonant and DIS region.  The detector technologies, MINERvA's segmented scintillator versus MicroBooNE's liquid Argon TPC, are also very different.  The success of the technique in both cases indicates that ML-based semantic segmentation can benefit $\pi^0$ reconstruction in a wide variety of neutrino detectors and beams.  
 

The behavior with $\Gamma$ score > 2 in Fig. \ref{fig:photon_score}b is consistent with an imperfect process, energy and spatial simulation being used to train the ML algorithm and/or used as the input to the score.
While this distribution is primarily $\gamma$ candidates, for the semi-exclusive pion production analysis it is primarily background and the discrepancy will be accounted for in the systematic uncertainty analysis.
The significant reduction in the background contamination shown in Table~\ref{table_0_700_60_200} should provide an improvement in the uncertainty analysis that should make up for the systematic associated with the ML algorithm which may be the source of this behavior.
The reconstruction includes a cut removing candidates with $\Gamma$ score > 2 and this is reflected in Fig. \ref{fig:blobsMass_BarbID}b.

While Fig. \ref{fig:photon_score}b shows a mismatch between data and simulation that increases linearly for a $\Gamma$ score greater than 2, the hypothesis that this arises due to bias coming from the semantic segmentation due to an unexpected mis-identification of hits caused by $\gamma$s can be disfavored because such hypothesis would suggest a correction that is more proportional to the number of events or linear in $\Gamma$ score.
While linearity is seen above a $\Gamma$ score of 2, it is not linear in the peak region between -2 and 2 where the vast majority of the $\gamma$ candidates reside.

The sample in the semi-inclusive analysis is significantly different than that of the semi-exclusive analysis due to the exclusion of a significant fraction of events with at least one charged pion.
That the same machine learning is able to provide a similarly strong filter for reconstruction of neutral pions from two $\gamma$s for both the semi-inclusive and semi-exclusive analyses with a similar match between data and simulation provides additional evidence that the machine learning algorithm is successfully filtering out energy from charged pions.

\subsection{Conclusions}
 
We have described various methods for neutral pion reconstruction: one is the baseline approach, one is a ML-based approach. The paper describes those approaches for a semi-exclusive analysis, with a different approach presented in the appendix. These measurements were done in a beam with $E_\nu$ around 6 GeV.  We have also described the development of a semantic segmentation-based filter on the hits used in the neutral pion reconstruction and demonstrated that it allows a significant reduction of mis-reconstructed $\gamma$s and thus mis-reconstructed neutral pions.
We expect significant increases in the final purity and efficiency of the analyses due to the improvement of the reconstruction purity from  approximately $70.7\pm0.9$\% to $89.3\pm0.7$\% in the peak region (60 to 200 MeV) of the invariant mass distributions (uncertainty is statistical).

Additionally, we have seen that the semantic segmentation-based filter reduces instances where  energy from other particles is included in the reconstructed pions.
This allows the energy to be fully available to other reconstructions which enables an improved estimate of the neutrino energy.
%
%
This approach thus provides a advantage over using multiple full event machine learning algorithms which may count features towards multiple quantities.
It also is an advantage over the baseline reconstruction which lacked the sophistication necessary to separate energy from a $\gamma$ from that of other particles within the reconstruction cone (see Section \ref{sec:emshowerreconstruction}).

This semantic segmentation based approach could be extended to include identifying the energy originating from protons, charged pions and more exotic hadrons.
Such an extension would be particularly interesting for semi-exclusive measurements of neutrino-induced meson production.
Semantic segmentation should be used to provide context to reconstructions in next generation experiments in High Energy Physics and should enable improved identification of $\nu_e$ appearance events.

%% file: acknowledgments.tex
\section*{Acknowledgments}
This document was prepared by members of the MINERvA Collaboration using the resources of the Fermi National Accelerator Laboratory (Fermilab), a U.S. Department of Energy, Office of Science, HEP User Facility. Fermilab is managed by Fermi Research Alliance, LLC (FRA), acting under Contract No. DE-AC02-07CH11359.
These resources included support for the \minerva construction project, and support
for construction also
was granted by the United States National Science Foundation under
Award No. PHY-0619727 and by the University of Rochester. Support for
participating scientists was provided by NSF and DOE (USA); by CAPES
and CNPq (Brazil); by CoNaCyT (Mexico); by Proyecto Basal FB 0821, CONICYT PIA ACT1413, Fondecyt 3170845 and 11130133 (Chile); 
by CONCYTEC (Consejo Nacional de Ciencia, Tecnolog\'ia e Innovaci\'on Tecnol\'ogica), DGI-PUCP (Direcci\'on de Gesti\'on de la Investigaci\'on  - Pontificia Universidad Cat\'olica del Peru), and VRI-UNI (Vice-Rectorate for Research of National University of Engineering) (Peru); NCN Opus Grant No. 2016/21/B/ST2/01092 (Poland); by Science and Technology Facilities Council (UK).  We thank the MINOS Collaboration for use of its near detector data. Finally, we thank the staff of
Fermilab for support of the beam line, the detector, and computing infrastructure.

%% file: sec_discussion.tex
\section{Appendix: Alternative Machine Learning models}
\label{sec:alternativesegmentations}

We also studied an alternative model where the neutron was included in the ``other'' class during the training of the ML algorithm.
In this study, the data was labeled with three classes and the segmentation was done by labeling each pixel as being one of two classes:  ``electromagnetic'' (EM-like) or ``non-electromagnetic" (non-EM-like).
The ``null'' labeled pixels were left unlabeled.
In this study the ML task is described as multi-class classification with 2 classes.
This model more closely follows the approach described in \cite{Adams:2018bvi}.

\begin{figure}[h!]
\centering
\includegraphics[width=15cm,height=9.5cm]{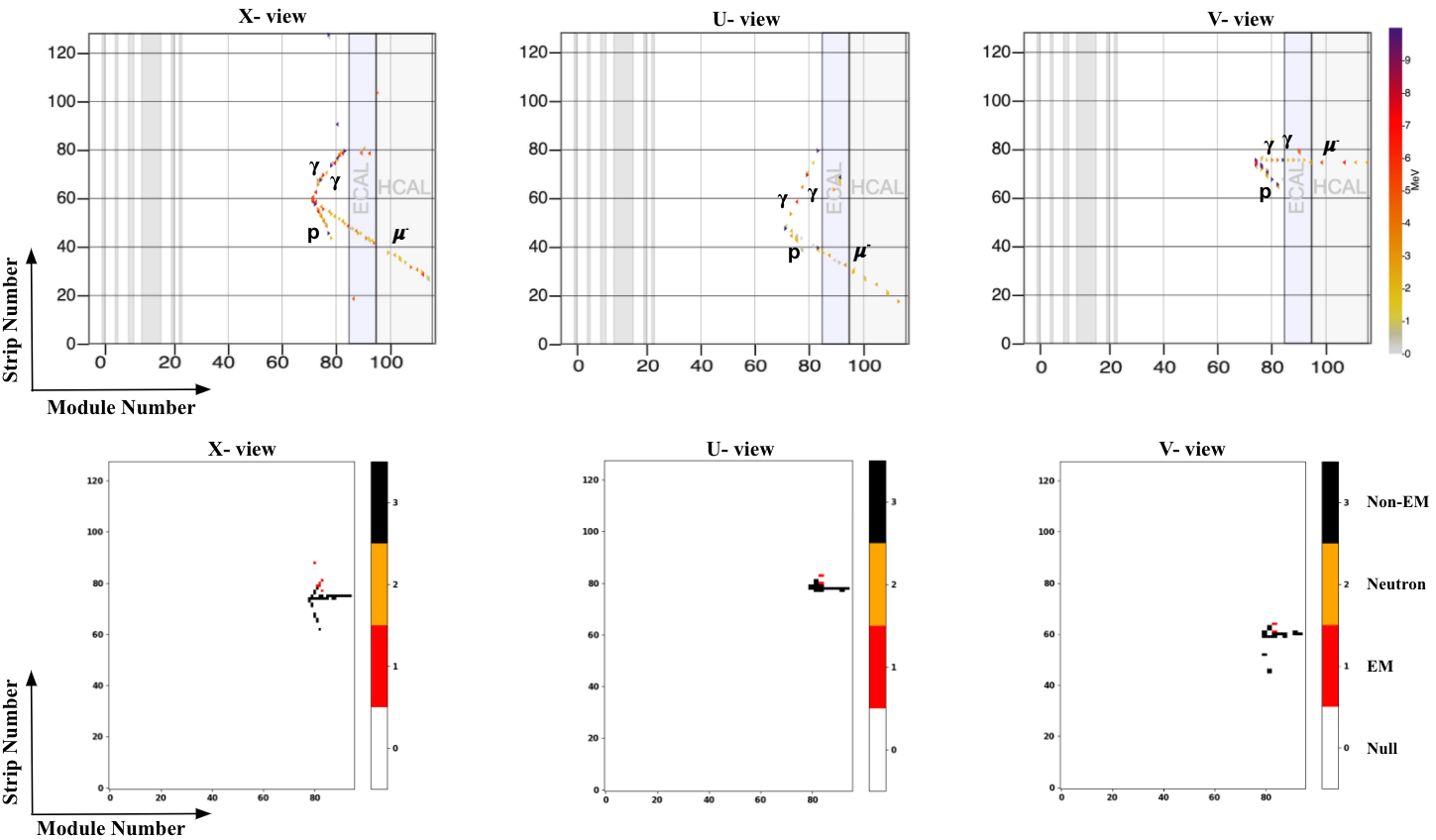}
\caption{
The lower panel represents the corresponding predicted images obtained by ML-based approach where the input images are formed by layered the three views together.
The color bar represents the integer values of each class.}
\label{fig:Eventdisplay3layer}
\end{figure}

In another alternative model, we also combined those three images into one by layering them \cite{8683736}.
We trained one network using those combined images and obtained the prediction for three views separately.
In Fig. \ref{fig:Eventdisplay3layer} we show result of the layered 4-class semantic segmentation model for one charged current neutral pion production event.

\begin{figure}[!htb]
\centering
\begin{subfigure}{0.4\textwidth}
  \centering
  \includegraphics[width=1.0\linewidth]{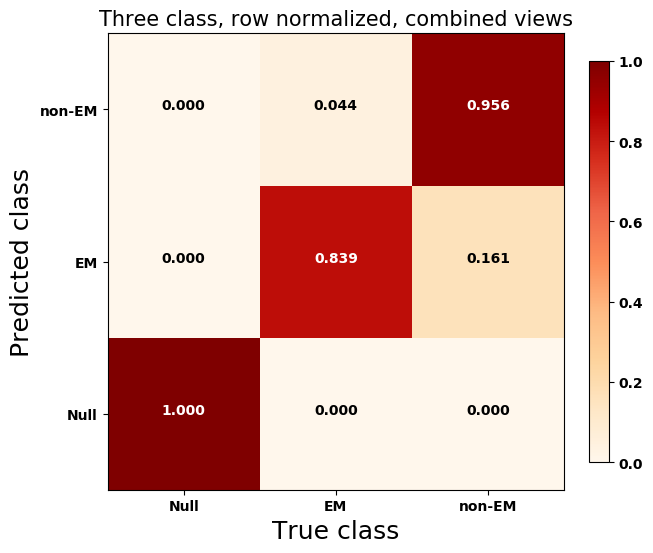}
  \caption{}
  \label{fig:row_norm3}
\end{subfigure}%
\begin{subfigure}{0.4\textwidth}
  \centering
  \includegraphics[width=1.0\linewidth]{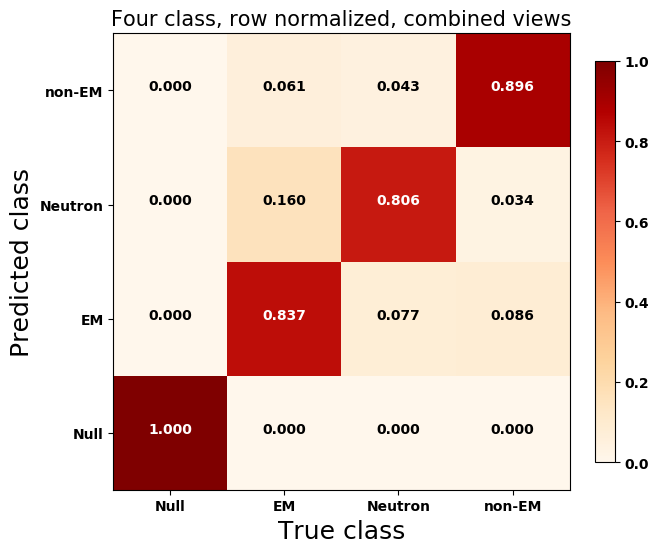}
  \caption{}
  \label{fig:row_norm_combined}
\end{subfigure}
\caption{ (a) Row-normalized confusion matrix with three labels: ``Null", EM-like, non-EM-like. 
Only X-views are used.
(b) Row Normalized confusion matrix with four labels: ``null'', EM-like, ``neutron'', non-EM-like shower where three views (X,U and V) are combined.
These confusion matrices are for a sample representative of the full training sample, including the nuclear target region.}
\label{fig:confusion_matrix}
\end{figure}

\begin{figure}[h!]
\centering
\includegraphics[width=15cm,height=9.5cm]{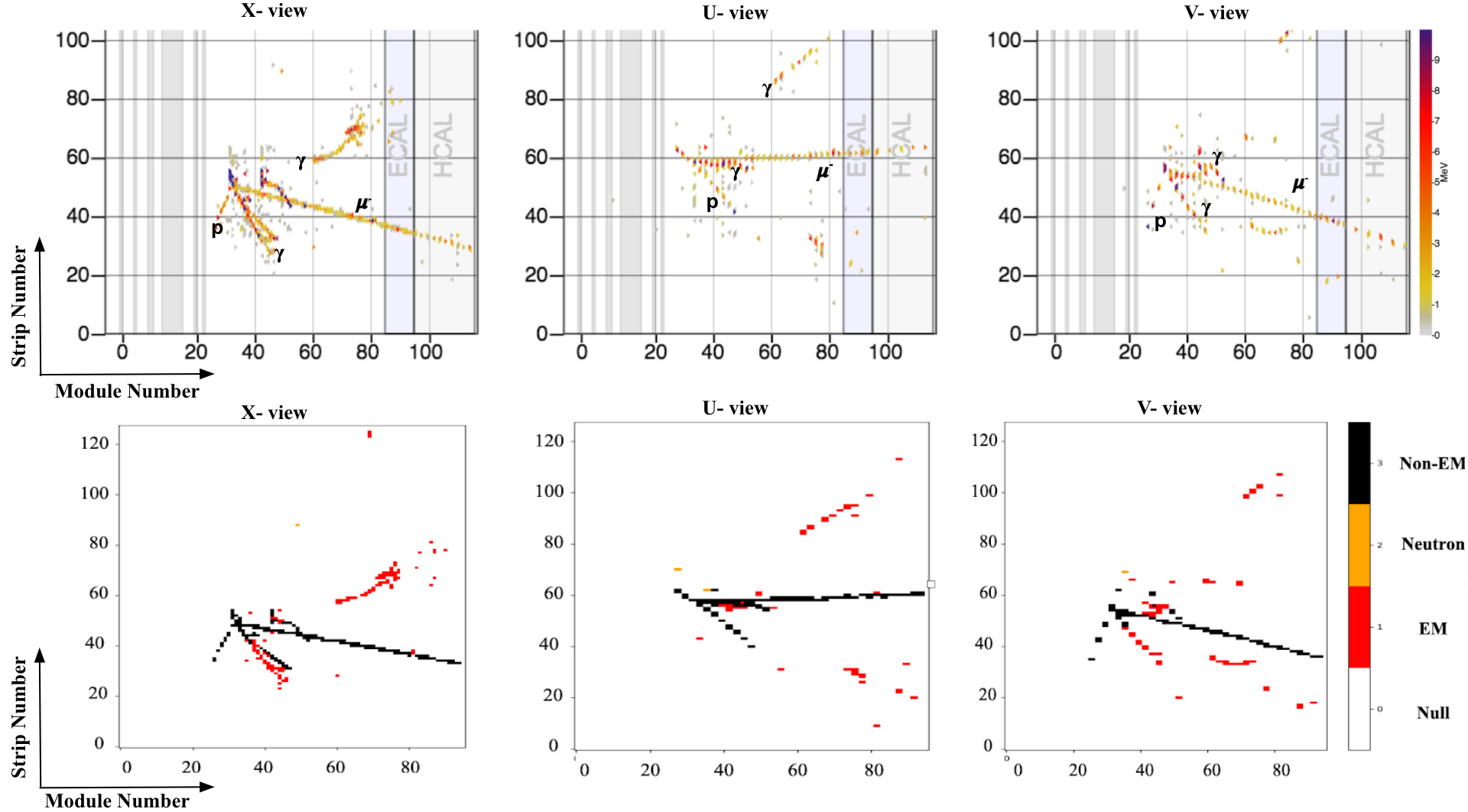}
\caption{Simulation of neutral pion production in \minerva for  DIS event with a neutron in the FS.  Here the three columns show the X, U and V-view of an event respectively.
The upper panel is the visualization of a MC event on web-based tool Arachne.
The lower panel represents the corresponding predicted images obtained by ML-based approach where the input images are formed by layered the three views together.
The color bar represents the integer values of each class.}
\label{fig:Event_neutron}
\end{figure}

Figure \ref{fig:row_norm3} shows the confusion matrix obtained by training over images with three labels such ``null", EM-like and non-EM-like.
On the other hand, Fig.\ref{fig:row_norm4} shows the confusion matrix with four labels such as ``null'', EM-like, ``neutron'' and non-EM-like.
The comparison between these two figures brings the conclusion that with four labels, the purity for EM-like hits improves by 5\% where as for non-EM-like hits the purity decreases by 3\%.
Figure \ref{fig:row_norm_combined} shows the confusion matrix where the images are formed by layered three views together ƒand are labeled into four classes such as ``null'', EM-like, ``neutron'' and non-EM-like.
Comparing Fig.\ref{fig:row_norm_combined} with Fig.\ref{fig:row_norm4} we see that the purity for EM-like hits and non-EM-like hits are decreased by 5\% and 3\% respectively for the combined image case whereas for ``neutron'' the purity is increased by 10\%. An example of an event with a neutron is provided in Fig. \ref{fig:Event_neutron}.

From our studies of alternative semantic segmentation models, including the ``neutron'' class improves the performance but layering did not, at least not to an X-view only application.
Additionally, the paucity of hits in U and V may complicate the application of the semantic segmentation due to the baseline reconstruction reconstructing candidate showers in the U and V-views before combining them into a candidate 3D shower.
A better application of semantic segmentation for U and V may be as a correction to the energy of the candidate $\gamma$ and not as a hit filter.

\newpage
\section{Appendix: Semi-inclusive analysis}
\label{sec:semiinclusive}

\begin{figure}[!htb]
\centering
\begin{subfigure}{0.5\textwidth}
  \centering
  \includegraphics[width=1.0\linewidth]{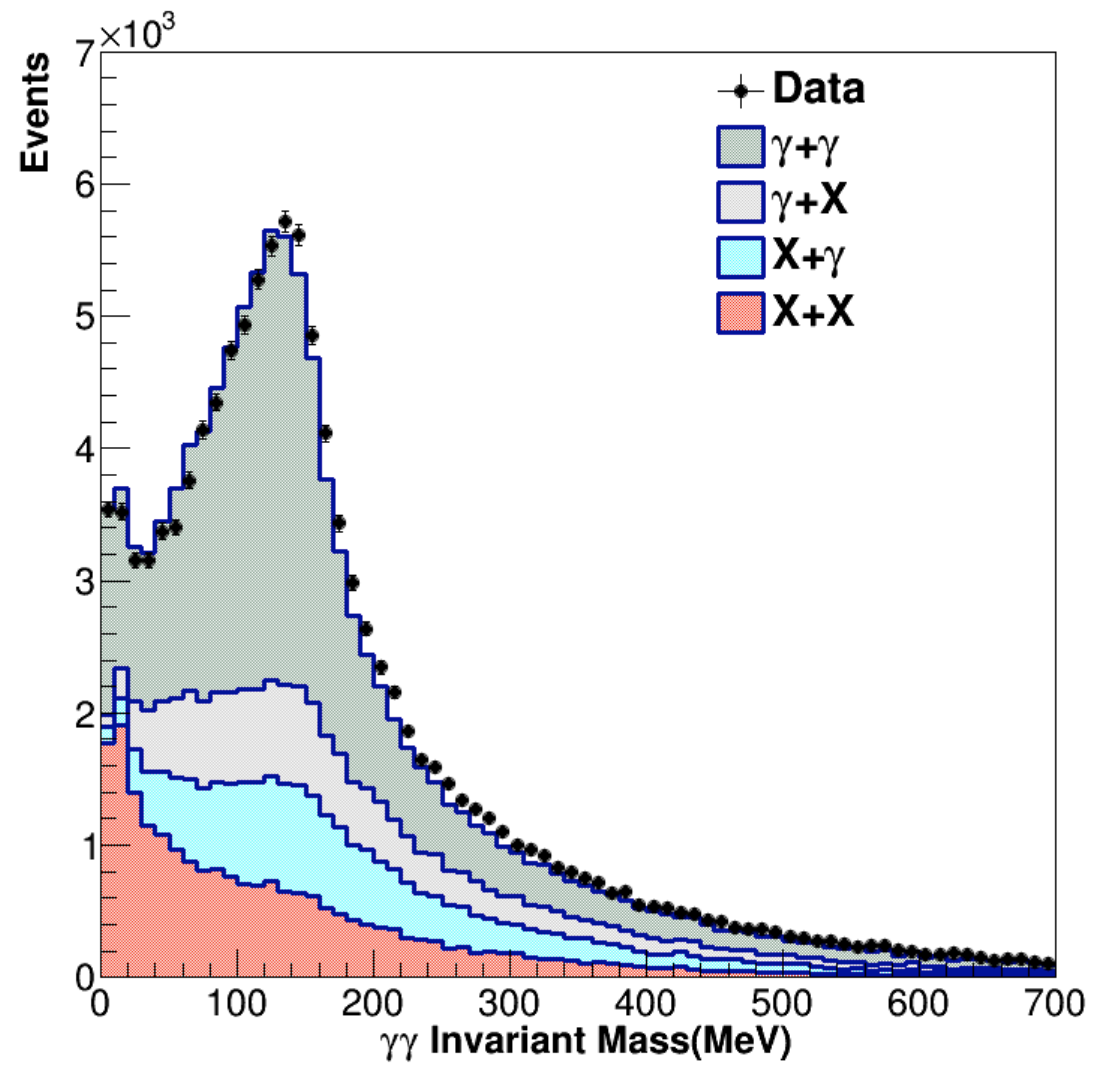}
  \caption{Baseline reconstruction of the neutral pion}
\end{subfigure}%
\begin{subfigure}{.5\textwidth}
  \centering
  \includegraphics[width=1.0\linewidth]{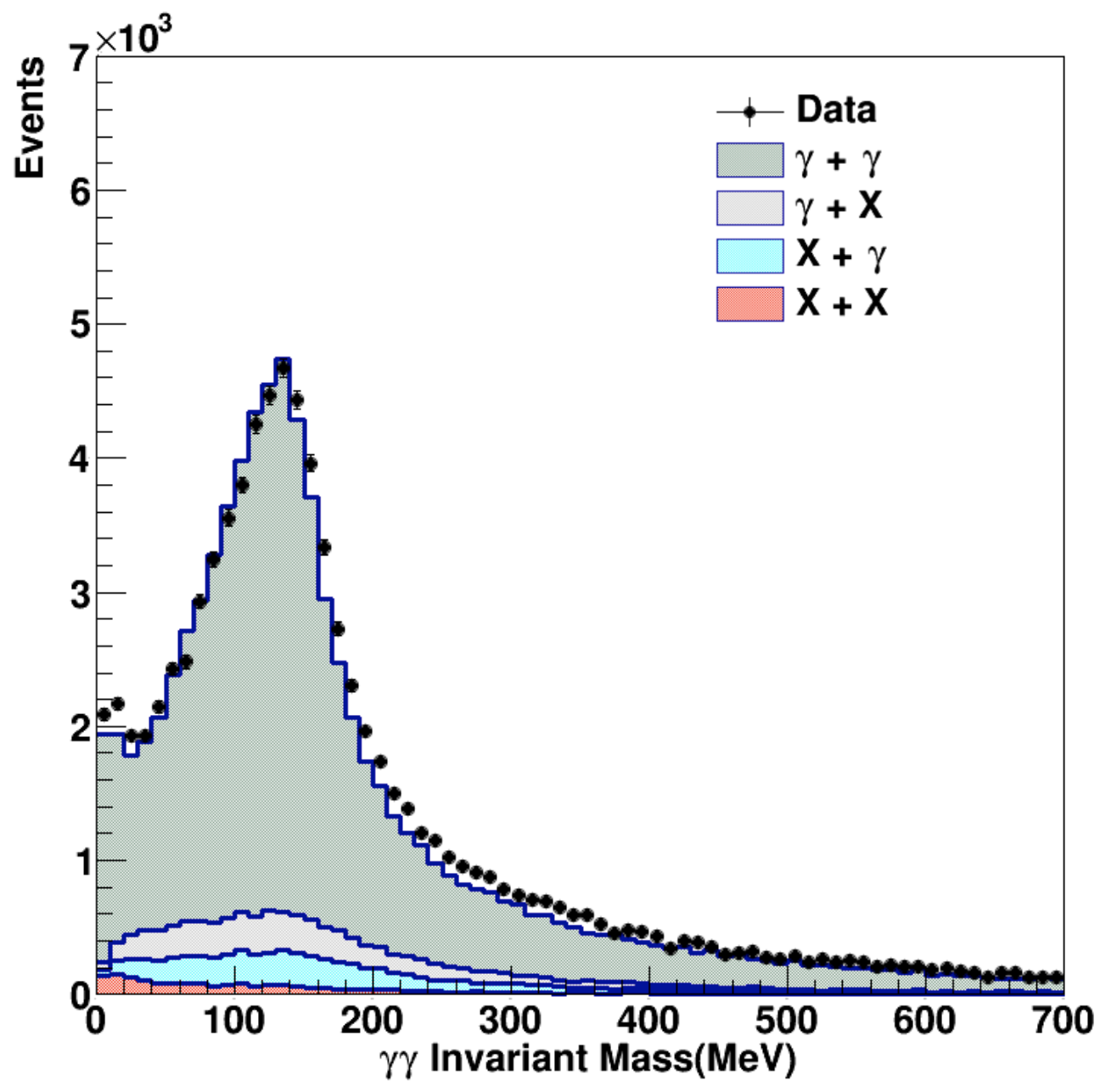}
  \caption{Neutral pion reconstruction with ML prediction}
\end{subfigure}
\caption{Neutral pion invariant mass distribution between $[0-700]$ MeV for the semi-inclusive analysis in data and simulation. The simulation is classified based on the true particle that contributed the greatest amount of energy to the primary and secondary $\gamma$ candidates. $X$ corresponds to the absence of a $\gamma$.  Note the significant improvement in the size of the $\gamma+\gamma$ category and decrease in all other categories when semantic segmentation provides context. The number of events in the simulation is normalized to equal that of the data.}
\label{fig:blobsMass_Roger}
\end{figure}

In Fig. \ref{fig:blobsMass_Roger} we see the invariant mass for both the baseline reconstruction and the reconstruction using the ML prediction for the semi-inclusive analysis.
The semi-inclusive analysis is similar to the semi-exclusive analysis, one exception being that the semi-exclusive analysis has additional filters designed to remove events with charged mesons, especially the positively charged pion ($\pi^+$).
By comparing Fig. \ref{fig:blobsMass_Roger} with Fig. \ref{fig:blobsMass_BarbID}, we see that the additional filters in the semi-exclusive analysis remove a significant number of the mis-reconstructions from non-$\gamma$s.
However, the reconstructions using semantic segmentation have a similar purity for both analyses.

Additionally, despite the very disparate samples in the two analyses, the application of semantic segmentation to the reconstruction does not significantly impact the agreement between data and simulation.
This provides of evidence for the success of our thesis, that semantic segmentation provides a strong hit filter which allows for the reconstruction of neutral pions independent of the existence of other energy within the detector which would normally have impacted the reconstruction.

%% file: main.bbl
\providecommand{\href}[2]{#2}\begingroup\raggedright\begin{thebibliography}{10}

\bibitem{Adamson:2015dkw}
P.~Adamson et~al., \emph{{The NuMI Neutrino Beam}},
  \href{https://doi.org/10.1016/j.nima.2015.08.063}{\emph{Nucl. Instrum. Meth.}
  {\bfseries A806} (2016) 279}
  [\href{https://arxiv.org/abs/1507.06690}{{\ttfamily 1507.06690}}].

\bibitem{Psihas:2020pby}
F.~Psihas, M.~Groh, C.~Tunnell and K.~Warburton, \emph{{A Review on Machine
  Learning for Neutrino Experiments}},
  \href{https://arxiv.org/abs/2008.01242}{{\ttfamily 2008.01242}}.

\bibitem{Guest_2018}
D.~Guest, K.~Cranmer and D.~Whiteson, \emph{Deep learning and its application
  to lhc physics},
  \href{https://doi.org/10.1146/annurev-nucl-101917-021019}{\emph{Annual Review
  of Nuclear and Particle Science} {\bfseries 68} (2018) 161–181}.

\bibitem{Qasim_2019}
S.~R. Qasim, J.~Kieseler, Y.~Iiyama and M.~Pierini, \emph{Learning
  representations of irregular particle-detector geometry with
  distance-weighted graph networks},
  \href{https://doi.org/10.1140/epjc/s10052-019-7113-9}{\emph{The European
  Physical Journal C} {\bfseries 79} (2019) }.

\bibitem{Andrews_2018}
M.~Andrews, M.~Paulini, S.~Gleyzer and B.~Poczos, \emph{End-to-end event
  classification of high-energy physics data},
  \href{https://doi.org/10.1088/1742-6596/1085/4/042022}{\emph{Journal of
  Physics: Conference Series} {\bfseries 1085} (2018) 042022}.

\bibitem{ad766625192f4650abd255078010c7cb}
D.~Belayneh, F.~Carminati, A.~Farbin, B.~Hooberman, G.~Khattak, M.~Liu et~al.,
  \emph{Calorimetry with deep learning: particle simulation and reconstruction
  for collider physics},
  \href{https://doi.org/10.1140/epjc/s10052-020-8251-9}{\emph{European Physical
  Journal C. Particles and Fields} {\bfseries 80} (2020) }.

\bibitem{Aurisano_2016}
A.~Aurisano, A.~Radovic, D.~Rocco, A.~Himmel, M.~Messier, E.~Niner et~al.,
  \emph{A convolutional neural network neutrino event classifier},
  \href{https://doi.org/10.1088/1748-0221/11/09/p09001}{\emph{Journal of
  Instrumentation} {\bfseries 11} (2016) P09001–P09001}.

\bibitem{Adamson_2016}
{\scshape MINOS} collaboration, \emph{{Measurement of single $\pi^0$ production
  by coherent neutral-current $\nu$ Fe interactions in the MINOS Near
  Detector}}, \href{https://doi.org/10.1103/PhysRevD.94.072006}{\emph{Phys.
  Rev. D} {\bfseries 94} (2016) 072006}
  [\href{https://arxiv.org/abs/1608.05702}{{\ttfamily 1608.05702}}].

\bibitem{Acero_2020}
M.~Acero, P.~Adamson, L.~Aliaga, T.~Alion, V.~Allakhverdian, N.~Anfimov et~al.,
  \emph{Measurement of neutrino-induced neutral-current coherent $\pi^{0}$
  production in the nova near detector},
  \href{https://doi.org/10.1103/physrevd.102.012004}{\emph{Physical Review D}
  {\bfseries 102} (2020) }.

\bibitem{Eberly_2015}
{\scshape MINERvA} collaboration, \emph{{Charged Pion Production in $\nu_\mu$
  Interactions on Hydrocarbon at $\langle E_{\nu}\rangle$= 4.0 GeV}},
  \href{https://doi.org/10.1103/PhysRevD.92.092008}{\emph{Phys. Rev. D}
  {\bfseries 92} (2015) 092008}
  [\href{https://arxiv.org/abs/1406.6415}{{\ttfamily 1406.6415}}].

\bibitem{lecun1995comparison}
Y.~LeCun, L.~Jackel, L.~Bottou, A.~Brunot, C.~Cortes, J.~Denker et~al.,
  \emph{Comparison of learning algorithms for handwritten digit recognition},
  in \emph{International conference on artificial neural networks}, vol.~60,
  pp.~53--60, Perth, Australia, 1995.

\bibitem{DBLP:journals/corr/RazavianASC14}
A.~S. Razavian, H.~Azizpour, J.~Sullivan and S.~Carlsson, \emph{{CNN} features
  off-the-shelf: an astounding baseline for recognition}, {\emph{CoRR}
  {\bfseries abs/1403.6382} (2014) }.

\bibitem{DBLP:journals/corr/LongSD14}
J.~Long, E.~Shelhamer and T.~Darrell, \emph{Fully convolutional networks for
  semantic segmentation}, {\emph{CoRR} {\bfseries abs/1411.4038} (2014) }
  [\href{https://arxiv.org/abs/1411.4038}{{\ttfamily 1411.4038}}].

\bibitem{PhysRevD.100.073005}
F.~Psihas, E.~Niner, M.~Groh, R.~Murphy, A.~Aurisano, A.~Himmel et~al.,
  \emph{Context-enriched identification of particles with a convolutional
  network for neutrino events},
  \href{https://doi.org/10.1103/PhysRevD.100.073005}{\emph{Phys. Rev. D}
  {\bfseries 100} (2019) 073005}.

\bibitem{DiFlorio:2018res}
{\scshape CMS} collaboration, \emph{{Convolutional Neural Network for Track
  Seed Filtering at the CMS High-Level Trigger}},
  \href{https://doi.org/10.1088/1742-6596/1085/4/042040}{\emph{J. Phys. Conf.
  Ser.} {\bfseries 1085} (2018) 042040}.

\bibitem{Abi:2018dnh}
{\scshape DUNE} collaboration, \emph{{The DUNE Far Detector Interim Design
  Report Volume 1: Physics, Technology and Strategies}},
  \href{https://arxiv.org/abs/1807.10334}{{\ttfamily 1807.10334}}.

\bibitem{Perdue:2012hg}
{\scshape MINERvA} collaboration, \emph{{The MINER$\nu$A Data Acquisition
  System and Infrastructure}},
  \href{https://doi.org/10.1016/j.nima.2012.08.024}{\emph{Nucl. Instrum. Meth.}
  {\bfseries A694} (2012) 179}
  [\href{https://arxiv.org/abs/1209.1120}{{\ttfamily 1209.1120}}].

\bibitem{Michael:2008bc}
{\scshape MINOS} collaboration, \emph{{The Magnetized steel and scintillator
  calorimeters of the MINOS experiment}},
  \href{https://doi.org/10.1016/j.nima.2008.08.003}{\emph{Nucl. Instrum. Meth.
  A} {\bfseries 596} (2008) 190}
  [\href{https://arxiv.org/abs/0805.3170}{{\ttfamily 0805.3170}}].

\bibitem{Aliaga:2013uqz}
{\scshape MINERvA} collaboration, \emph{{Design, Calibration, and Performance
  of the MINERvA Detector}},
  \href{https://doi.org/10.1016/j.nima.2013.12.053}{\emph{Nucl. Instrum. Meth.
  A} {\bfseries 743} (2014) 130}
  [\href{https://arxiv.org/abs/1305.5199}{{\ttfamily 1305.5199}}].

\bibitem{Fields_2013}
L.~Fields, J.~Chvojka, L.~Aliaga, O.~Altinok, B.~Baldin, A.~Baumbaugh et~al.,
  \emph{Measurement of muon antineutrino quasielastic scattering on a
  hydrocarbon target at $\langle e_\nu \rangle \sim 3.5$ gev},
  \href{https://doi.org/10.1103/physrevlett.111.022501}{\emph{Physical Review
  Letters} {\bfseries 111} (2013) }.

\bibitem{Fiorentini_2013}
G.~A. Fiorentini, D.~W. Schmitz, P.~A. Rodrigues, L.~Aliaga, O.~Altinok,
  B.~Baldin et~al., \emph{Measurement of muon neutrino quasielastic scattering
  on a hydrocarbon target at $\langle e_\nu \rangle \sim 3.5$gev},
  \href{https://doi.org/10.1103/physrevlett.111.022502}{\emph{Physical Review
  Letters} {\bfseries 111} (2013) }.

\bibitem{Tice_2014}
B.~Tice, M.~Datta, J.~Mousseau, L.~Aliaga, O.~Altinok, M.~Barrios~Sazo et~al.,
  \emph{Measurement of ratios of $\nu_{\mu}$ charged-current cross sections on
  c, fe, and pb to ch at neutrino energies 2–20 gev},
  \href{https://doi.org/10.1103/physrevlett.112.231801}{\emph{Physical Review
  Letters} {\bfseries 112} (2014) }.

\bibitem{Wolcott_2016}
{\scshape MINERvA} collaboration, \emph{{Measurement of electron neutrino
  quasielastic and quasielasticlike scattering on hydrocarbon at $\langle
  E_{\nu} \rangle $ = 3.6 GeV}},
  \href{https://doi.org/10.1103/PhysRevLett.116.081802}{\emph{Phys. Rev. Lett.}
  {\bfseries 116} (2016) 081802}
  [\href{https://arxiv.org/abs/1509.05729}{{\ttfamily 1509.05729}}].

\bibitem{Park_2016}
J.~Park, L.~Aliaga, O.~Altinok, L.~Bellantoni, A.~Bercellie, M.~Betancourt
  et~al., \emph{Measurement of neutrino flux from neutrino-electron elastic
  scattering}, \href{https://doi.org/10.1103/physrevd.93.112007}{\emph{Physical
  Review D} {\bfseries 93} (2016) }.

\bibitem{Ruterbories_2019}
D.~Ruterbories, K.~Hurtado, J.~Osta, F.~Akbar, L.~Aliaga, D.~Andrade et~al.,
  \emph{Measurement of quasielastic-like neutrino scattering at $\langle e_\nu
  \rangle \sim 3.5 $gev on a hydrocarbon target},
  \href{https://doi.org/10.1103/physrevd.99.012004}{\emph{Physical Review D}
  {\bfseries 99} (2019) }.

\bibitem{Carneiro_2020}
M.~Carneiro, D.~Ruterbories, Z.~Ahmad~Dar, F.~Akbar, D.~Andrade, M.~Ascencio
  et~al., \emph{High-statistics measurement of neutrino quasielasticlike
  scattering at 6 gev on a hydrocarbon target},
  \href{https://doi.org/10.1103/physrevlett.124.121801}{\emph{Physical Review
  Letters} {\bfseries 124} (2020) }.

\bibitem{Altinok:2017xua}
{\scshape MINERvA} collaboration, \emph{{Measurement of $\nu_{\mu}$
  charged-current single $\pi^{0}$ production on hydrocarbon in the few-GeV
  region using MINERvA}},
  \href{https://doi.org/10.1103/PhysRevD.96.072003}{\emph{Phys. Rev. D}
  {\bfseries 96} (2017) 072003}
  [\href{https://arxiv.org/abs/1708.03723}{{\ttfamily 1708.03723}}].

\bibitem{Coplowe:2020yea}
{\scshape MINERvA} collaboration, \emph{{Probing nuclear effects with
  neutrino-induced charged-current neutral pion production}},
  \href{https://doi.org/10.1103/PhysRevD.102.072007}{\emph{Phys. Rev. D}
  {\bfseries 102} (2020) 072007}
  [\href{https://arxiv.org/abs/2002.05812}{{\ttfamily 2002.05812}}].

\bibitem{Aliaga:2015wva}
{\scshape MINERvA} collaboration, \emph{{Single Neutral Pion Production by
  Charged-Current $\bar{\nu}_\mu$ Interactions on Hydrocarbon at $\langle E_\nu
  \rangle = $3.6 GeV}},
  \href{https://doi.org/10.1016/j.physletb.2015.07.039}{\emph{Phys. Lett. B}
  {\bfseries 749} (2015) 130}
  [\href{https://arxiv.org/abs/1503.02107}{{\ttfamily 1503.02107}}].

\bibitem{Wolcott:2016hws}
{\scshape MINERvA} collaboration, \emph{{Evidence for Neutral-Current
  Diffractive $\pi^0$ Production from Hydrogen in Neutrino Interactions on
  Hydrocarbon}},
  \href{https://doi.org/10.1103/PhysRevLett.117.111801}{\emph{Phys. Rev. Lett.}
  {\bfseries 117} (2016) 111801}
  [\href{https://arxiv.org/abs/1604.01728}{{\ttfamily 1604.01728}}].

\bibitem{Tagg:2011wk}
{\scshape MINERvA} collaboration, \emph{{Arachne - A web-based event viewer for
  MINERvA}}, \href{https://doi.org/10.1016/j.nima.2012.01.059}{\emph{Nucl.
  Instrum. Meth. A} {\bfseries 676} (2012) 44}
  [\href{https://arxiv.org/abs/1111.5315}{{\ttfamily 1111.5315}}].

\bibitem{genie1}
C.~Andreopoulos et~al., \emph{{The GENIE Neutrino Monte Carlo Generator}},
  \href{https://doi.org/10.1016/j.nima.2009.12.009}{\emph{Nucl. Instrum. Meth.}
  {\bfseries A614} (2010) 87}
  [\href{https://arxiv.org/abs/0905.2517}{{\ttfamily 0905.2517}}].

\bibitem{Agostinelli:2002hh}
{\scshape GEANT4} collaboration, \emph{{GEANT4--a simulation toolkit}},
  \href{https://doi.org/10.1016/S0168-9002(03)01368-8}{\emph{Nucl. Instrum.
  Meth. A} {\bfseries 506} (2003) 250}.

\bibitem{REIN198179}
D.~Rein and L.~M. Sehgal, \emph{Neutrino-excitation of baryon resonances and
  single pion production},
  \href{https://doi.org/https://doi.org/10.1016/0003-4916(81)90242-6}{\emph{Annals
  of Physics} {\bfseries 133} (1981) 79 }.

\bibitem{bodek2003higher}
A.~BODEK and U.~YANG, \emph{Higher twist, $\xi$ $\omega$ scaling, and effective
  lo pdfs for lepton scattering in the few gev region}, {\emph{Journal of
  physics. G. Nuclear and particle physics} {\bfseries 29} (2003) 1899}.

\bibitem{PhysRevD.24.1400}
A.~Bodek and J.~L. Ritchie, \emph{Further studies of fermi-motion effects in
  lepton scattering from nuclear targets},
  \href{https://doi.org/10.1103/PhysRevD.24.1400}{\emph{Phys. Rev. D}
  {\bfseries 24} (1981) 1400}.

\bibitem{doi:10.1063/1.3661588}
S.~A. Dytman and A.~S. Meyer, \emph{Final state interactions in genie},
  \href{https://doi.org/10.1063/1.3661588}{\emph{AIP Conference Proceedings}
  {\bfseries 1405} (2011) 213}
  [\href{https://arxiv.org/abs/https://aip.scitation.org/doi/pdf/10.1063/1.3661588}{{\ttfamily
  https://aip.scitation.org/doi/pdf/10.1063/1.3661588}}].

\bibitem{Perdue:2018ihs}
{\scshape MINERvA} collaboration, \emph{{Reducing model bias in a deep learning
  classifier using domain adversarial neural networks in the MINERvA
  experiment}},
  \href{https://doi.org/10.1088/1748-0221/13/11/P11020}{\emph{JINST} {\bfseries
  13} (2018) P11020} [\href{https://arxiv.org/abs/1808.08332}{{\ttfamily
  1808.08332}}].

\bibitem{Stowell_2019}
P.~Stowell, L.~Pickering, C.~Wilkinson, C.~Wret, F.~Akbar, D.~Andrade et~al.,
  \emph{Tuning the genie pion production model with minerva data},
  \href{https://doi.org/10.1103/physrevd.100.072005}{\emph{Physical Review D}
  {\bfseries 100} (2019) }.

\bibitem{larcv}
``{LarCV}.'' \url{https://github.com/DeepLearnPhysics/larcv2}, 2017.

\bibitem{Brun:1997pa}
R.~Brun and F.~Rademakers, \emph{{ROOT: An object oriented data analysis
  framework}},
  \href{https://doi.org/10.1016/S0168-9002(97)00048-X}{\emph{Nucl.Instrum.Meth.}
  {\bfseries A389} (1997) 81}.

\bibitem{uresnetcite}
``{U-ResNet}.'' \url{https://github.com/DeepLearnPhysics/u-resnet}, 2018.

\bibitem{Adams:2018bvi}
{\scshape MicroBooNE} collaboration, \emph{{Deep neural network for pixel-level
  electromagnetic particle identification in the MicroBooNE liquid argon time
  projection chamber}},
  \href{https://doi.org/10.1103/PhysRevD.99.092001}{\emph{Phys. Rev. D}
  {\bfseries 99} (2019) 092001}
  [\href{https://arxiv.org/abs/1808.07269}{{\ttfamily 1808.07269}}].

\bibitem{microboone}
``{MicroBooNE Collaboration}.'' \url{https://microboone-exp.fnal.gov/}.

\bibitem{DBLP:journals/corr/RonnebergerFB15}
O.~Ronneberger, P.~Fischer and T.~Brox, \emph{U-net: Convolutional networks for
  biomedical image segmentation}, {\emph{CoRR} {\bfseries abs/1505.04597}
  (2015) } [\href{https://arxiv.org/abs/1505.04597}{{\ttfamily 1505.04597}}].

\bibitem{DBLP:journals/corr/HeZRS15}
K.~He, X.~Zhang, S.~Ren and J.~Sun, \emph{Deep residual learning for image
  recognition}, {\emph{CoRR} {\bfseries abs/1512.03385} (2015) }
  [\href{https://arxiv.org/abs/1512.03385}{{\ttfamily 1512.03385}}].

\bibitem{726791}
Y.~{Lecun}, L.~{Bottou}, Y.~{Bengio} and P.~{Haffner}, \emph{Gradient-based
  learning applied to document recognition},
  \href{https://doi.org/10.1109/5.726791}{\emph{Proceedings of the IEEE}
  {\bfseries 86} (1998) 2278}.

\bibitem{DBLP:journals/corr/IoffeS15}
S.~Ioffe and C.~Szegedy, \emph{Batch normalization: Accelerating deep network
  training by reducing internal covariate shift}, {\emph{CoRR} {\bfseries
  abs/1502.03167} (2015) } [\href{https://arxiv.org/abs/1502.03167}{{\ttfamily
  1502.03167}}].

\bibitem{Valencia:2019mkf}
{\scshape MINERvA} collaboration, \emph{{Constraint of the MINER$\nu$A medium
  energy neutrino flux using neutrino-electron elastic scattering}},
  \href{https://doi.org/10.1103/PhysRevD.100.092001}{\emph{Phys. Rev. D}
  {\bfseries 100} (2019) 092001}
  [\href{https://arxiv.org/abs/1906.00111}{{\ttfamily 1906.00111}}].

\bibitem{Adams:2019law}
{\scshape MicroBooNE} collaboration, \emph{{Reconstruction and Measurement of
  $\mathcal{O}$(100) MeV Energy Electromagnetic Activity from $\pi^0
  \rightarrow \gamma\gamma$ Decays in the MicroBooNE LArTPC}},
  \href{https://doi.org/10.1088/1748-0221/15/02/P02007}{\emph{JINST} {\bfseries
  15} (2020) P02007} [\href{https://arxiv.org/abs/1910.02166}{{\ttfamily
  1910.02166}}].

\bibitem{8683736}
L.~{Song}, F.~{Chen}, S.~R. {Young}, C.~D. {Schuman}, G.~{Perdue} and T.~E.
  {Potok}, \emph{Deep learning for vertex reconstruction of neutrino-nucleus
  interaction events with combined energy and time data},  in \emph{ICASSP 2019
  - 2019 IEEE International Conference on Acoustics, Speech and Signal
  Processing (ICASSP)}, pp.~3882--3886, 2019.

\end{thebibliography}\endgroup
